%% file: 24RoughVolAAPRevision02.tex
\documentclass[11pt,paper=a4]{article}
 \usepackage{indentfirst, latexsym,bm}
 \usepackage{amsmath}
 \usepackage{pifont}
 \usepackage{amsfonts}
 \usepackage{mathrsfs}
 \usepackage{array}
 \usepackage{multirow} 
 \usepackage{graphicx}
 \usepackage{subfigure}
 \usepackage{picinpar}
 \usepackage{setspace}
 \usepackage[textsize=footnotesize]{todonotes}
 \RequirePackage[colorlinks,citecolor=blue,urlcolor=blue,linkcolor=blue]{hyperref}
 \usepackage[top=2.5cm,bottom=2.5cm, outer=2.5cm, inner=2.5cm]{geometry}

 \RequirePackage[numbers]{natbib}
 \usepackage{enumitem} 
 
 \usepackage{booktabs}
 \usepackage{authblk}
 
 \usepackage{threeparttable}
 \usepackage{mathrsfs,amsfonts,amsmath}
 
 \usepackage{color}
 
 \makeatletter
 \def\namedlabel#1#2{\begingroup
 	#2%
 	\def\@currentlabel{#2}%
 	\phantomsection\label{#1}\endgroup
 }
 \makeatother

 \newcommand{\red}{\color{red}}
 \newcommand{\blue}{\color{blue}}
  \newcommand{\magenta}{\color{magenta}}
 
 \newcommand\email[1]{\rm\href{mailto:#1}{ \nolinkurl{#1}}}

 \renewcommand{\theequation}{\arabic{section}.\arabic{equation}}
 
 \newtheorem{theorem}{Theorem}[section]
 \newtheorem{definition}[theorem]{Definition}
 \newtheorem{lemma}[theorem]{Lemma}
 \newtheorem{corollary}[theorem]{Corollary}
 \newtheorem{proposition}[theorem]{Proposition}
 \newtheorem{remark}[theorem]{Remark}
 \newtheorem{condition}[theorem]{Condition}
 \newtheorem{example}{Example}[section]

 \def\red{\color{red}}\def\blue{\color{blue}}
 
 \def\blemma{\begin{lemma}}\def\elemma{\end{lemma}}
 \def\bproposition{\begin{proposition}}\def\eproposition{\end{proposition}}
 \def\ttheorem{\begin{theorem}}\def\etheorem{\end{theorem}}
 \def\bcorollary{\begin{corollary}}\def\ecorollary{\end{corollary}}
 \def\bremark{\begin{remark}}\def\eremark{\end{remark}}
 \def\bcondition{\begin{condition}}\def\econdition{\end{condition}}

 \newtheorem{assumption}[theorem]{Assumption}

 \def\benumerate{\begin{enumerate}}\def\eenumerate{\end{enumerate}}
 \def\bitemize{\begin{itemize}}\def\eitemize{\end{itemize}}

 \def\beqlb{\begin{eqnarray}}\def\eeqlb{\end{eqnarray}}
 \def\beqnn{\begin{eqnarray*}}\def\eeqnn{\end{eqnarray*}}
 \def\ar{\!\!\!&}

 \def\proof{\noindent{\it Proof.~~}}\def\qed{\hfill$\Box$\medskip}

\begin{document}

 \title{\bf  Convergence of Heavy-Tailed Hawkes Processes and the Microstructure of Rough Volatility\thanks{We thank Peter Bank and Masaaki Fukasawa for valuable feedback.}}

 \author{Ulrich Horst\footnote{Department of Mathematics and School of Business and Economics,  Humboldt-Universit\"at zu Berlin, Unter den Linden 6, 10099 Berlin; email: horst@math.hu-berlin.de. Horst gratefully acknowledges financial support by the German Research Foundation trough CRCTRR 388 {\sl Rough Analysis, Stochastic Dynamics and Related Fields}, Project B02.}\quad\ \  
 Wei Xu\footnote{School of Mathematics and Statistics, Beijing Institute of Technology, No. 5, South Street, Zhongguancun, Haidian District, 100081 Beijing; email: xuwei.math@gmail.com}
 \quad\  and \quad Rouyi Zhang\footnote{Department of Mathematics, Humboldt-Universit\"at zu Berlin, Unter den Linden 6, 10099 Berlin; email: rouyi.zhang@hu-berlin.de}
 }
   \maketitle

 \begin{abstract}
   We establish the weak convergence of the intensity of a nearly-unstable Hawkes process with heavy-tailed kernel. 
   Our result is used to derive a scaling limit for a financial market model where orders to buy or sell an asset arrive according to a Hawkes process with power-law kernel. 
   After suitable rescaling the price-volatility process converges weakly to a rough Heston model. 
   Our convergence result is stronger than previously established ones that have either focused on light-tailed kernels or the convergence of integrated volatility process. The key is to establish the tightness of the family of rescaled volatility processes. 
   This is achieved by introducing a new methods to establish the $C$-tightness of c\`adl\`ag processes based on the classical Kolmogorov-Chentsov tightness criterion for continuous processes. 
 	
 	\medskip
 	\smallskip
 	
 	\noindent \textbf{\textit{MSC2020 subject classifications.}} 
 	 Primary 60G55, 60G22;  secondary 60F05
 	 
 	  \smallskip
 	 
 \end{abstract}

 \input{Introduction}

\input{BenchmarkModel}

 \input{ConvergenceVol}

 \appendix
 
 \renewcommand{\theequation}{A.\arabic{equation}}
 \input{MittagLeffler}
 
 \bibliographystyle{plain}
 \bibliography{Reference}

 \end{document}

%% file: Introduction.tex
 \section{Introduction and overview}
 First introduced by Hawkes in \cite{Hawkes1971a,Hawkes1971b} to model cross-dependencies between earthquakes and their aftershocks, Hawkes processes have long become a powerful tool to model a variety of phenomena in the sciences, humanities, economics and finance. 

 A {Hawkes process} is a random point process $\{N(t):t\geq 0\}$ that models self-exciting arrivals of random events. In such settings, events arrive at random points in time $\tau_1 < \tau_2 < \tau_3 < \cdots$ according to an  \textit{intensity} process $\{V(t):t\geq 0\}$ that is usually of the form
 \beqlb\label{HawkesDensity}
 V(t):= \mu(t)+ \sum_{0<\tau_i<t} \phi(t-\tau_i) = \mu(t) + \int_{(0,t)} \phi(t-s)N(ds), \quad t \geq 0,
 \eeqlb
 where the \textit{immigration density} $\mu(\cdot)$ captures the arrival of exogenous events and the \textit{kernel} $\phi(\cdot)$ captures the self-exciting impact of past events on the arrivals of future events. 

Applications of Hawkes processes in finance include intraday transaction dynamics \cite{Bauwens2009,Bowsher2007}, asset price and limit order book dynamics \cite{BacryDelattreHoffmannMuzy2013, HorstXu2019}, financial contagion \cite{SahaliaCacho-DiazLaeven2015,Giesecke2011,JorionZhang2009} and - in particular - stochastic volatility modeling \cite{ElEuchFukasawaRosenbaum2018,HorstXu2022,JaissonRosenbaum2015,JaissonRosenbaum2016,RosenbaumTomas2021}.  We consider a stochastic volatility model where orders to buy or sell an asset arrive according to a Hawkes process with a {\sl heavy-tailed kernel} of the form
 \beqnn
	\phi(t) = \alpha  \sigma  \cdot \big(1+\sigma \cdot  t\big)^{-\alpha - 1} \quad \mbox{for some constants} \quad \alpha \in (1/2,1) \ \mbox{ and }\ \sigma>0
 \eeqnn
 and prove the weak convergence of a sequence of suitably rescaled volatility process to a fractional diffusion and the joint convergence of the price-volatility process to a rough Heston model. 

 Our model is strongly inspired by the ones studied in \cite{ElEuchFukasawaRosenbaum2018,JaissonRosenbaum2016,RosenbaumTomas2021} but our convergence results are much more refined. 
While the aforementioned works focused on the {\sl integrated} volatility processes we establish a weak convergence result for the volatility process itself. This provides a strong microstructure foundation for a class of rough volatility models. 

 Microstructure models of financial markets have been extensively studied in the financial mathematics and economics literature in the last decades. This literature often provides economic foundations for the use of specific market models by linking important model assumptions and features to investor and/or order arrival characteristics. As pointed out by O’Hara in her influential book {\sl Market Microstructure Theory} \cite{OHara}, it was Garman’s 1976 paper \cite{Garman} that inaugurated the explicit study of market microstructure. He argued that ``market agents can be treated as a statistical ensemble [and that] their market activities [can be] depicted as the stochastic generation of market orders according to a Poisson process''. In this point of view, one models right away the aggregate order flow rather than characterizing agents’ investment decisions as solutions to individual utility maximization problems. 
 Garman's approach was later extended beyond the benchmark case of Poisson dynamics by many authors including \cite{Foellmer94, FoellmerSchweizer, Horst2005} to provide microstructure foundation for the emergence of financial bubbles and, more recently, microstructure foundations of stochastic volatility models \cite{ElEuchFukasawaRosenbaum2018,HorstXu2022,JaissonRosenbaum2015,JaissonRosenbaum2016}.

 \subsection{Hawkes process and (rough) volatility}

 Hawkes processes with light-tailed kernels have been used in Horst and Xu \cite{HorstXu2022} to establish scaling limits for a class of continuous-time stochastic volatility models with self-exciting jump dynamics. 
 Many of the existing jump diffusion stochastic volatility models including the classical Heston model \cite{Heston1993}, the Heston model with jumps \cite{Bates1996,Bates2019,Pan2002}, the OU-type volatility model \cite{Barndorff-NielsenShephard2001}, the multi-factor model with self-exciting volatility spikes \cite{Bates2019} and the alpha Heston model \cite{JiaoMaScottiZhou2021} were obtained as scaling limits under different scaling regimes. 

 Nearly unstable Hawkes processes with light-tailed kernels were first analyzed by Jaisson and Rosenbaum \cite{JaissonRosenbaum2015}.  
 They proved the weak convergence of the rescaled intensity to a Feller diffusion - also known as CIR-model in finance - and the convergence of the rescaled point process to the integrated diffusion; their result was extended to multi-variate processes in \cite{Xu2021}. 
 Under a heavy-tailed condition on the kernel, Jaisson and Rosenbaum \cite{JaissonRosenbaum2016} later considered the weak convergence of the rescaled point process to the integral of a rough fractional diffusion; a corresponding convergence result for the characteristic function of the rescaled Hawkes process has been considered in \cite{ElEuchRosenbaum2019b}. 
 Analogous scaling limits in the multivariate case were established in  \cite{ElEuchFukasawaRosenbaum2018,RosenbaumTomas2021}. 

 The aforementioned works provide microstructural foundations for large classes of stochastic volatility models, including the standard Heston model. The Heston model assumes that the volatility process follows a Brownian semi-martingale. 
 However, the analysis in \cite{GatheralJaissonRosenbaum2018} suggests that historical volatility time series are much rougher than those of Brownian martingales and that log-volatility is better modeled by a fractional Brownian motion with a Hurst parameter $H < 1/2$. 
 
 This observation spurred substantial research on the properties and the microstructural foundations of rough volatility models.  
 For instance, a rough Heston model of the form
 \beqnn
 dS(t) \ar=\ar S(t)\sqrt{V(t)}dW(t) , \cr
 \ar\ar \cr
 V(t) \ar=\ar \int_0^t  \frac{(t-s)^{\alpha-1}}{\Gamma(\alpha)} \cdot b\Big(\theta - V(s) \Big) ds
 +\int_0^t \frac{(t-s)^{\alpha-1}}{\Gamma(\alpha)}\cdot \gamma \sqrt{V(s)}dB(s),
 \eeqnn
 has been introduced \cite{ElEuchFukasawaRosenbaum2018, ElEuchRosenbaum2019b}. 
 The rough Heston model is an affine Volterra process \cite{JaberLarssonPulido2019} and admits a semi-explicit representation of the characteristic function in terms of a fractional Riccati equation.  
 Other popular rough volatility models include the quadratic rough Heston model \cite{RosenbaumZhang2021}, the (mixed) rough Bergomi model \cite{JacquierMartiniMuguruza2018,LacombeMuguruzaStone2021} and the rough SABR model \cite{FukasawaGatheral2022}.  

 Although estimating the precise degree of roughness of volatility is subtle and challenging empirically (see \cite{ChongHoffmannLiuRosenbaumSzymanski2023a, ChongHoffmannLiuRosenbaumSzymanski2023b,ContDas2022,Fukasawa2021,HanSchied23} and references therein for a detailed discussion) the use of rough volatility models is by now an established paradigm for modeling equity markets. 
 Rough volatility provides excellent fits to market data; in particular, it reproduces very well the behaviour of the implied volatility surface, in particular the at-the-money skew that it is often observed in practice as shown in, e.g. \cite{BayerFrizGatheral2016}. 
 At the same time, the lack of Markovianity and regularity of the sample paths leads to significant challenges for sample path simulation and option pricing. 
 Markovian approximations and asymptotic expansions as established in, e.g.~\cite{BayerBreneis2023a,BayerBreneis2023b,FrizWagenhofer2023,Fukasawa2011} or the recently introduced rough PDE approach for local stochastic volatility models \cite{BankBayerFrizPelizzari2023} provide partial remedies to these practical problems.

 \subsection{Our contribution}

 Our analysis is inspired by and complements the earlier work on microstructure foundations of rough volatility models by Rosenbaum and co-workers, especially \cite{ElEuchFukasawaRosenbaum2018,JaissonRosenbaum2016,RosenbaumTomas2021}. 
 We first introduce a family  of order driven financial market models where orders to buy and sell an asset arrive according to a Hawkes process, and each order changes the logarithmic price by a random amount. 
 In other words, our logarithmic price process is driven by a Hawkes point measure as introduced in \cite{HorstXu2022}; price processes driven by multi-variate Hawkes processes are contained as a special case. 

 Subsequently, we consider a family of rescaled volatility/intensity processes $\{V^{(n)}\}_{n \geq 1}$ where the intensity of order arrivals tends to infinity, the impact of an individual order on the price price process tends to zero and the average number of child orders triggered by each mother order tends to one. 
 We prove the weak convergence of the sequence $\{V^{(n)}\}_{n \geq 1}$ in the usual Skorokhod space $\mathbf{D}(\mathbb{R}_+;\mathbb{R})$ of all $\mathbb{R}$-valued c\`{a}dl\`{a}g functions on $\mathbb{R}_+$ endowed with the Skorokhod topology to a fractional diffusion. 
 
 Previous work on microstrucual foundation of rough volatility models focused on the weak convergence of the integrated volatility processes
 \beqnn
	\mathcal{I}_{V^{(n)}}(t) := \int_0^t V^{(n)}(s) \, ds, \quad t \geq 0.
 \eeqnn

The weak convergence of the sequence of integrated volatility processes does in general {\sl not} imply the convergence of the volatility processes. Even if the sequence of integrated processes would converge, one cannot obtain the limit of the sequence $\{V^{(n)}\}_{n \geq 1}$ by differentiating the limit of the sequence $\{\mathcal{I}_{V^{(n)}}\}_{n \geq 1}$ unless the $C$-tightness of the former sequence has been established.\footnote{ A tight sequence of processes is called $C$-tight if any weak accumulation point is continuous; see Definition~3.25 in \cite[p.351]{JacodShiryaev2003}.} This calls for more refined convergence results for the volatility process that we establish in this paper. 

 As already argued in \cite{JaissonRosenbaum2015} the main challenge is to prove the $C$-tightness of the sequence of rescaled volatility processes; 
 the increments of the volatility process  do not satisfy the standard moment condition that is usually required to establish the $C$-tightness of a sequence of stochastic processes.  To overcome this challenge we introduce a novel technique to verify the $C$-tightness of a sequence {\sl c\`adl\`ag} processes based on the classical Kolmogorov-Chentsov tightness criterion for {\sl continuous} processes. 
%
%

A family of {c\`adl\`ag} stochastic processes is $C$-tight if it can be approximated in probability by $C$-tight processes. 
Our key observation is that a family of c\`adl\`ag processes $\{X^{(n)}\}_{n \geq 1}$ admits a $C$-tight approximation by piecewise linear processes defined on the time grids 
\[
	\big\{ k/n^\theta : k=0, ..., [Tn^\theta] \big\} \quad \mbox{for some} \quad \theta > 2, 
\]	
if (i) the initial states satisfy a uniform moment condition; (ii) the jumps sizes at the grid points converge to zero uniformly in probability, 
\beqlb 
	\displaystyle\sup_{k=0,1,\cdots,[Tn^\theta]} \sup_{h\in[0,1/n^{\theta}]} \big|\Delta_h X^{(n)}(k/n^\theta) \big|  \overset{\rm p}\to 0
\eeqlb 
as $n \to \infty$; (iii) the increments at arbitrary time points satisfy the polynomial moment condition 
 	\beqlb 
 	 \sup_{t\in [0,T]}	\mathbf{E}\Big[\big|\Delta_h X^{(n)}(t)\big|^{p}\Big]\leq 
 		C\cdot \sum_{i=1}^m \frac{h^{b_i}}{n^{a_i}}
 	\eeqlb 
for all $h \in (0,1)$, some integer $m$ and suitable constants $(a_i,b_i)$, $i=1, ..., m$. Our non-standard moment condition resembles the classical Kolmogorov-Chentsov tightness criterion for continuous processes but it is much weaker and applies to general c\`adl\`ag processes.\footnote{One way to prove the Kolmogorov-Chentsov tightness condition is by approximation on dyadic time grids. We work with coarser time yet very specific grids $\{t^{(n)}_i\}_{i=i}^{M^{(n)}}$.} 

 The uniform moment condition on the initial states of stochastic integral terms is not difficult to establish. Establishing the convergence of jumps to zero along the chosen sequence of grid points is more subtle. 
 The key is to increase the number of grid-points at the correct rate, and then to apply a suitable moment estimate for stochastic integrals driven by Poisson random measures. 
 The added difficulty in our setting is that one of the upper integral boundaries is specified by a stochastic process (the volatility process). A powerful moment estimate for such integrals has recently been established in \cite{Xu2021b}. 
 Applying this same result again, we then prove that the discontinuities of the stochastic integral term satisfies the non-standard moment condition.  

With the $C$-tightness of the sequence  $\{V^{(n)}\}_{n \geq 1}$ in hand, we then proceed to prove the weak convergence of the rescaled volatility processes. An application of Skorokhod’s representation theorem shows that the weak convergence of the rescaled volatility processes implies the weak convergence of the sequence $\{\mathcal{I}_{V^{(n)}}\}_{n \geq 1}$ - as pointed out above, the converse is {\sl not} true in general. We then utilize the weak convergence of the integrated processes along a subsequence to prove the uniqueness of the weak accumulation points of the sequence $\{V^{(n)}\}_{n \geq 1}$ and hence the convergence of the rescaled volatility processes. 

 We strongly emphasize that our analysis utilizes the convergence of the integrated volatility process only to identify the weak limits of the sequence $\{V^{(n)}\}_{n \geq 1}$. Furthermore, 
our approach strongly hinges on the $C$-tightness of the sequence of the rescaled volatility processes. 
 Without the $C$-tightness of the volatility processes we can neither prove the convergence of the integrated processes, nor utilize the limit of the integrated processes to identify the limiting volatility process.   
 
 Having established the weak convergence of the volatility process, the joint convergence of the rescaled price-volatility process to a rough Heston-type model is then easily obtained. As a byproduct we also obtain the weak convergence of the rescaled Hawkes process.

The remainder of this paper is organized as follows. In Section \ref{Section02}, we introduce our benchmark model financial market model and the state the main results of this paper. All proofs are given in Section \ref{Sec3}.

 \medskip
 
 \textit{\textbf{Notation.}} We denote by $[x]$ the integer part of  the real number $x\in\mathbb{R}$ and put $\mathbb{R}_+=[0,\infty)$. 

 For two real-valued functions $f,g$ on $\mathbb{R}_+$, we define their convolution $f*g$ by 
 \beqnn
 f*g(t):= \int_0^t f(t-s)g(s)ds = \int_0^t f(s)g(t-s)ds,\quad t\geq 0,
 \eeqnn 
 and write $f^{*n}$ for the $n$-th convolution of $f$.
 For any $p\in(0,\infty]$, we denote by $L^{p}(\mathbb{R};\mathbb{R})$ the space of measurable functions $f$ on $\mathbb{R}$  the space of all functions $f$ on $\mathbb{R}$ that satisfy $\|f\|_{L^p}^p:=\int_\mathbb{R} |f(s)|^p ds<\infty$ and by  $L^p_{\rm loc}(\mathbb{R};\mathbb{R})$  the space of all functions $f$ on $\mathbb{R}$ such that $\int_{|s|\leq T} |f(s)|^p ds<\infty$ for any $T\geq 0$.

 Almost sure convergence,  convergence in distribution and convergence in probability is denoted by   $\overset{\rm a.s.}\to$, $\overset{\rm d}\to$ and $\overset{\rm p}\to$, respectively. 
 We write  $\overset{\rm a.s.}=$, $\overset{\rm d}=$ and $\overset{\rm p}=$ to denote almost sure equality, equality in distribution and equality in probability.
 
   Throughout this paper, we denote by $C$ a generic constant that may vary from line to line.

%% file: BenchmarkModel.tex
 \section{Heavy-tailed Hawkes market model} \label{Section02}
 \setcounter{equation}{0}
 
 In this section we introduce a benchmark asset price model for which we drive a scaling limit in a later section. As pointed out above, the model is similar to the one studied in \cite{ElEuchFukasawaRosenbaum2018,JaissonRosenbaum2016,RosenbaumTomas2021} but we derive more refined convergence results. 
  
We assume throughout that all random variables and stochastic processes are defined on a common probability space $(\Omega, \mathscr{F} , \mathbf{P})$ endowed with a filtration $\{\mathscr{F}_t : t \geq 0\}$ that satisfies the usual hypotheses. The convergence concept for stochastic processes we use will be weak convergence in the space $\mathbf{C}(\mathbb{R}_+;\mathbb{R}^d)$ of all $\mathbb{R}^d$-valued continuous functions on $\mathbb{R}_+$ endowed with the uniform topology or in the space $\mathbf{D}(\mathbb{R}_+;\mathbb{R}^d)$ of all $\mathbb{R}^d$-valued c\`{a}dl\`{a}g functions on $\mathbb{R}_+$ endowed with the Skorokhod topology; see \cite{Billingsley1999,JacodShiryaev2003}.

 \subsection{The benchmark model}
 
 We consider an order-driven model where asset prices are driven by incoming orders to buy or sell the asset. The order arrivals times are described by an increasing sequence of $(\mathscr{F}_t)$-adapted random times $\{\tau_k  \}_{k\geq 1}$. The impact of each order on the price is described by an independent sequence of i.i.d.~$\mathbb R$-valued random variables $\{\xi_k \}_{k\geq 1}$ with distribution $\nu(du)$.  In terms of these sequences we define the random point measure
 \beqnn
 N(ds, du) := \sum_{k=1}^\infty {\bf 1}_{\{\tau_k \in ds, \xi_k \in du\}}
 \eeqnn
 on $(0,\infty) \times \mathbb R$ and assume that the logarithmic price process $\{P(t) : t \geq 0\}$ satisfies the dynamics 
 \beqlb\label{eqn.Price}
 P(t) \ar=\ar  P(0)+   \sum_{\tau_k\leq t}  \xi_k = P(0)+  \int_0^t \int_{\mathbb R}   u \, N(ds,du),
 \eeqlb
 where $P(0)$ is the initial price at time zero. 
 
 \begin{remark}
 	The special case $\nu(du) = \frac{1}{2} \cdot \delta_{-1}(du) +  \frac{1}{2} \cdot\delta_{1}(du)$ corresponds to the special case where each order increases or decreases the log price by one tick with equal probability. Here $\delta_x(du)$ is the Dirac measure at point $x$. 
 \end{remark}
 
 We assume that $N(ds,du)$ is a \textit{marked Hawkes point measure}. More precisely, we assume that the embedded point process 
 \beqnn
 N(t) := N\big((0,t],\mathbb{R}\big),\quad t\geq 0,
 \eeqnn
 is a Hawkes process with intensity process
 \beqlb\label{eqn.Vol}
 	V(t):= \Lambda(t)+ \mu  + \sum_{k=1}^{N(t)}\zeta  \cdot \phi(t-\tau_k) =\Lambda(t)+ \mu  + \int_0^t  \zeta \cdot \phi(t-s)dN(s), \quad t \geq 0. 
 \eeqlb
 The function $\Lambda\in \mathbf{D}(\mathbb{R}_+;\mathbb{R}_+)$ represents the combined impact of all the events that arrived prior to time zero on future arrivals;  the positive constant intensity $\mu $ describes the arrival rate of ``exogenous'' orders; the kernel {\color{blue} $\phi \in \mathbf{D}(\mathbb{R}_+;\mathbb{R}_+) \cap L^1(\mathbb{R}_+;\mathbb{R}_+)$} specifies the self-exciting impact of past order arrivals on future arrivals; the positive constant $\zeta$ measures the impact of each child order on the overall order arrival dynamics. We call the vectors
 \beqnn
 \big( \Lambda,\mu ,\zeta,\phi,\nu \big) 
 \quad \mbox{and} \quad 
 \big(\Lambda,\mu ,\zeta,\phi\big)
 \eeqnn
 the \textit{characteristic} of the point measure $N(ds, du)$ and the embedded Hawkes process $\{N(t) : t \geq 0\}$, respectively. 
 
 \begin{remark} \label{rem:representation}
 An explicit construction of the marked Hawkes point measure $N(ds,du)$ using Crump-Mode-Jagers branching processes with immigration is given in \cite{Xu2021}. 
%
%
%
		The embedded Hawkes process $\{N(t) : t \geq 0\}$ on the positive half real line with intensity $V$ given by \eqref{eqn.Vol} has been analyzed in many works, including~\cite{BacryDelattreHoffmannMuzy2013, GaoZhu2018c}. In particular, it has been shown that the process is non-explosive if $\|\phi\|_{L^1}<\infty$. 
		Hawkes processes fron the whole real line are extensively analyzed in the textbook \cite{Bremaud2020}.

  \end{remark}
  
 Accounting for the empirically well-documented long-range dependencies in order arrivals we consider a power-law kernel of the form  
 \beqlb\label{eqn.Kernel}
  \phi(t) = \alpha  \sigma  \cdot \big(1+\sigma  \cdot t\big)^{-\alpha - 1} 
  \quad \mbox{for some constants} \quad \alpha \in (1/2,1) \ \mbox{ and }\ \sigma>0.
 \eeqlb
 The function $\phi$ is a probability density function on $\mathbb{R}_+$ with tail-distribution 
 \beqlb\label{eqn.TailvarPhi}
 \overline\varPhi(t):= \int_t^\infty \phi(s) ds = \big(1+\sigma  \cdot t \big)^{-\alpha},\quad t\geq 0.
 \eeqlb
 
 Associated with the kernel $\phi$ we define the \textit{resolvent} $R$ by the unique solution of the resolvent equation 
 \beqlb\label{eqn.Resolvent}
 	R(t)= \zeta\cdot\phi(t)+  \zeta\cdot\phi*R(t),\quad t\geq 0.
 \eeqlb  
  The resolvent will play an important role when analyzing the dynamics of the volatility process.  Under our assumptions on the kernel $\phi$ it follows from, e.g.~Theorem 3.1 in \cite[p.42]{GripenbergLondenStaffans1990} or Lemma~3 in  \cite{BacryDelattreHoffmannMuzy2013} that the resolvent equation and hence the resolvent is well defined; since our kernel is continuous, the resolvent is continuous, too. \footnote{Lemma 3 in \cite{BacryDelattreHoffmannMuzy2013} states that for a locally bounded function $h$ on $\mathbb{R}_+$, there exists a unique locally bounded solution to the equation $f(t)=h(t)+ \zeta\cdot\phi *f(t) $ which is given by $f(t) = h(t)+ R*h(t)$.} 
 
 \begin{remark}
 	In our benchmark model each order triggers an average number 
   \beqnn
 	\zeta \cdot \|\phi\|_{L_1} = \zeta   
 	\eeqnn
 	of child orders. Following \cite{ElEuchFukasawaRosenbaum2018,JaissonRosenbaum2016,RosenbaumTomas2021}  we assume that $\zeta < 1$ and then consider a family of rescaled models where the number of child orders tends to one.
 \end{remark}
 
 To give give a more convenient representation of the price-volatility dynamics we apply Theorem~7.4 in \cite[p.93]{IkedaWatanabe1989}
			\footnote{The theorem states that given a measurable space $(\boldsymbol{X},\mathscr{B}(\boldsymbol{X}))$ and a random point measure $N_p(dt,d\boldsymbol{x})$ on $(0,\infty)\times \boldsymbol{X}$ with intensity $dt\,q(t,d\boldsymbol{x})$ 				where $q(t,d\boldsymbol{x})$ is a progressive measure-valued process on $(\boldsymbol{X},\mathscr{B}(\boldsymbol{X}))$, then if there exists a $\sigma$-finite measure $m(d\boldsymbol{z})$ on a standard measurable space $(\boldsymbol{Z},			\mathscr{B}(\boldsymbol{Z}))$ and a predictable process $\theta(t,\boldsymbol{z}) : \mathbb{R}_+\times \boldsymbol{Z} \mapsto \boldsymbol{X}$ such that $$m(\{\boldsymbol{z}: \theta(t,\boldsymbol{z}) \in E \}) \overset{\rm a.s.}= q(t,E), \quad E \in 			\mathscr{B}(\boldsymbol{X}),$$ 
			then on an extension of the original probability space we can define a time-homogeneous Poisson random measure $N_q(dt,d\boldsymbol{z})$ on $(0,\infty)\times  \mathbf{Z}$ with intensity $dt\, m(d\boldsymbol{z})$ such that  
 			$$N_p((0,t],E) = \int_0^t \int_{\boldsymbol{Z}} \mathbf{1}_{\{ \theta(s,\boldsymbol{z}) \in E \}} N_q(ds, d\boldsymbol{z}),\quad E \in \mathscr{B}(\boldsymbol{X}). $$} 
		 with $\boldsymbol{X}=\mathbb{R}$, $\boldsymbol{Z}=\mathbb{R}\times\mathbb{R}_+$, $\boldsymbol{x}=x$, $\boldsymbol{z}=(u,z)$ and 
 		$$
 		m(du,dz)=\nu(du)dz,\quad q(t,dx)=V(t-) \nu(dx),\quad 
 		\theta\big(t,(u,z)\big)= u\cdot \mathbf{1}_{\{0<z\leq V(t-)\}} +{\red \infty \cdot \mathbf{1}_{\{z> V(t-)\}} }.
 		$$	
 		It then follows that on an extension of the original probability space we can define a Poisson random measure $N_0(ds,du,dz)$ on $(0,\infty)\times \mathbb R \times (0,\infty)$ with intensity measure $ ds \, \nu (du) \, dz$ such that our Hawkes process can be conveniently represented as  
 		\beqnn
 		N(t)= \int_0^t \int_\mathbb{R} \int_0^{V(s-)} N_0(ds,du,dz) ,\quad t\geq 0, 
 		\eeqnn
 		and $$N(ds,du):= N_0\big(ds,du,(0,V(s-)]\big)$$ is a realization of our  marked Hawkes point measure. As a result, we can represent the price-volatility process as
 \begin{equation}
 \begin{split}
 P(t) & = P(0)+\int_0^t \int_{\mathbb{R}} \int_0^{V(s-)}  u\, N_0(ds, du, dz), \quad t \geq 0 ,\cr
 V(t) & = \Lambda(t)+ \mu  +  \int_0^t  \int_0^{V(s-)} \zeta \cdot \phi(t-s) N_1 (ds, dz), \quad t \geq 0,
 \end{split}
 \end{equation}
 where $N_1(ds, dz) := N_0(ds, \mathbb{R}, dz)$. It is well known that the compensated Poisson random measures 
 \beqnn
 \widetilde{N}_0(ds,du,dz):= N_0(ds,du,dz)-  ds \, \nu (du) \, dz \quad \mbox{and} \quad
 \widetilde{N}_1(ds,dz):= N_1(ds,dz)-  ds\,dz
 \eeqnn
 are two $(\mathscr{F}_t)$-martingale measures. 
 In terms of the compensated measures,
 the price-volatility process can be represented as the unique solution to the integral equation 
 \begin{equation} 
 	\begin{split}
 		P(t) &= P(0) +  \int_{\mathbb R} u \nu (du) \cdot \int_0^t V(s)ds + \int_0^t \int_{\mathbb R} \int_0^{V(s-)}   u \, \widetilde{N}_0(ds, du, dz), \quad t \geq 0 ,\\ 
 		V(t) &=  \zeta\cdot\phi*V(t)+ \Lambda(t)  + \mu   + \int_0^t \int_0^{V(s-)}  \zeta\cdot\phi(t-s) \widetilde{N}_1(ds,dz) , \quad t \geq 0. 
 	\end{split}
 \end{equation} 
 As in the proof of Proposition~2.1 in \cite{JaissonRosenbaum2015}, we can apply Lemma~3
 in \cite{BacryDelattreHoffmannMuzy2013} as well the equation \eqref{eqn.Resolvent} and the stochastic Fubini's theorem (see  \cite[Theorem~2.6]{Walsh1986}) to write the price-volatility process as the unique solution to the integral equation 
 \begin{equation} \label{eqn.PoissonVol}
 \begin{split}
  P(t) &= P(0) +  \int_{\mathbb R} u \nu (du) \cdot \int_0^t V(s)ds + \int_0^t \int_{\mathbb R} \int_0^{V(s-)}   u \, \widetilde{N}_0(ds, du, dz), \quad t \geq 0 ,\\ 
  V(t) &=  \Lambda(t) + \Lambda* R(t) + \mu + \mu\int_0^t R(s)ds  + \int_0^t \int_0^{V(s-)}  R(t-s) \widetilde{N}_1(ds,dz), \quad t \geq 0. 
 \end{split}
 \end{equation}  
 \subsection{Scaling limits}
 
 In what follows we introduce a sequence of rescaled models where the intensity of order arrivals tends to infinity, the impact of an individual order tends to zero and the average number of child orders tends to one. The dynamics of the price-volatility process $\big(P_n, V_n \big)$ in $n$-th market model  is defined in terms of an underlying marked Hawkes point measure $N_n(ds,du)$ with characteristic $$\big(\Lambda_n,\mu_n , \zeta_n,\phi,\nu_n \big),$$ akin to the dynamics \eqref{eqn.Price} and \eqref{eqn.Vol}. The corresponding Hawkess process that specifies the order arrival dynamics is denoted  $\{N_n(t):t\geq 0\}$; the corresponding Poisson random measures are denoted by, respectively 
 \[
 	{N}_{n,0}(ds,du,dz) \quad \mbox{and} \quad {N}_{n,1}(ds,dz). 
\]

 \subsubsection{The volatility process} 

 The main challenge is to establish the weak convergence of the rescaled volatility process on which we focus in this subsection. 
 Specifically, we consider the
%
%
 sequence of rescaled volatility processes defined by 
 \beqnn
 V^{(n)}(t) := \frac{V_n(nt)}{n^{2 \alpha - 1}}, \quad t \geq 0. 
 \eeqnn
 Applying a change of variables to (\ref{eqn.PoissonVol}), the rescaled process satisfies
 \beqlb \label{rescaledV}
 	V^{(n)}(t)
 	\ar=\ar \frac{\Lambda_n(nt)}{n^{2\alpha-1}}+
 	\frac{R_n*\Lambda_n (nt)}{n^{2\alpha-1}}  + \frac{\mu_n}{n^{2\alpha-1}}+
 	\frac{\mu_n }{n^{2\alpha-1}}\int_0^{nt}R_n(s)ds \cr
 	\ar\ar  + \int_0^t \int_0^{V^{(n)}(s-)}\frac{R_n\big(n(t-s)\big)}{ n^{2\alpha-1}}\widetilde{N}_{n,1}(n\cdot ds ,n^{2\alpha-1}\cdot dz), \quad t \geq 0,
 \eeqlb
 where $R_n$ is the resolvent defined as in (\ref{eqn.Resolvent}) with $\zeta=\zeta_n$, and
 \beqnn
   \widetilde{N}_{n,1}(n\cdot ds ,n^{2\alpha-1}\cdot dz):= N_{n,1}(n\cdot ds ,n^{2\alpha-1}\cdot dz)- n^{2\alpha} \cdot  ds \, dz.   
 \eeqnn
 
 We assume throughout that the model parameters $( \Lambda_n,\mu_n,\zeta_n )$ satisfy the following condition. In particular, we assume that the expected number of child orders increases to one and that the impact of orders that arrived prior to time zero is slowly decaying. 
 
 \begin{assumption}\label{main.Condition}
  Assume that the following hold.
  \begin{enumerate}
   \item[(1)] On average, each order triggers less than one child order, that is, $\zeta_n<1$ for each $n\geq 1$.
 		
 \item[(2)] The function $\Lambda_n$ satisfies $\Lambda_n(t) =V_{n,0}\cdot \overline\varPhi(t)  $ with $ V_{n,0} \in \mathbb{R}_+$ for each $n\geq 1$ and $t\geq 0$.
 
 \item[(3)] There exist three constants $V_*(0) \in \mathbb{R}$, $a\geq 0$ and $b>0$ such that as $n\to\infty$, 
 \beqlb
 \frac{V_{n,0}}{n^{2\alpha-1}}\to V_*(0) ,\quad \frac{\mu_n}{n^{\alpha-1}}	\to a , \quad 
 n^\alpha(1-\zeta_n)\to  b .
 \eeqlb
 
  \end{enumerate}
 \end{assumption}
 
  \begin{remark}
 	Let us briefly comment on our initial state condition.
 	It is well known that a standard 
 	Hawkes process $N$ defined on the whole real line with intensity 
 	\beqnn
 	\lambda(t) = \mu + \int_{-\infty}^t \phi(t-s) dN(s)
 	=\mu + \int_{-\infty}^0 \phi(t-s) dN(s) +\int_0^t \phi(t-s) dN(s),\quad t\in \mathbb{R}
 	\eeqnn
 	 is stationary if each event triggers less than one child event on average, i.e., $\|\phi\|_{L^1}<1$. 
 	The first integral on the right side of the second equality is the impact  of all past events prior to time zero at time $t>0$ with mean impact 
 	\begin{align*}
 	 \mathbf{E}\Big[ \int_{-\infty}^0 \phi(t-s) dN(s)  \Big] & =\frac{\mu}{1-\|\phi\|_{L^1}}\int_{-\infty}^0 \phi(t-s) ds \\
	 & = \frac{\mu}{1-\|\phi\|_{L^1}}\int_{t}^\infty \phi(s) ds \\
	 & = \frac{\mu}{1-\|\phi\|_{L^1}} \cdot \overline\varPhi(t).
 	\end{align*}
 	Here $\frac{\mu}{1-\|\phi\|_{L^1}}$ is the mean  residual impact of all  events that happened prior to time zero at time zero. It decreases according to the function $\overline\varPhi$. 
 	It is hence reasonable to assume that conditioned on the residual impact $V_{n,0}$, the impact of the events that happened prior to time zero on the arrival of future events at any time $t > 0$ equals 
 	\beqnn
 	V_{n,0} \cdot \int_{-\infty}^0 \phi(t-s)ds = 	V_{n,0} \cdot \int_t^{\infty} \phi(s)ds =V_{n,0}\cdot \overline\varPhi(t).
 	\eeqnn 
	This explains our choice of $\Lambda_n$.
	It is worth noting that our initial condition is different from that in \cite[Definition 2.3]{ElEuchRosenbaum2019b} where for technical reasons the authors had to assume that 
    \beqnn
   {\magenta \Lambda_n =} V_{n,0}\cdot \Big( \overline\varPhi(t) -n^{\alpha-1} \int_0^t \phi(s)ds\Big). 
    \eeqnn
 \end{remark}
  
 We are now ready to state our first main result. It states that our sequence of rescaled volatility processes converges weakly to a unique limiting process; the proof is given in Section \ref{Sec.proofs}. 
 
 To formulate our result we denote by $F^{\alpha,\gamma}$ and $f^{\alpha,\gamma}$ the Mittag-Leffler probability distribution and density function\footnote{We recall the definition and basic properties of the Mittag-Leffler function and the Mittag-Leffler density function in Appendix~\ref{App.ML}.} respectively with parameters $(\alpha, \gamma)$, where $$\gamma := \frac{b  \sigma^\alpha }{ \Gamma(1-\alpha)}$$ and $\Gamma(\cdot)$ is the gamma function. 
 
 \begin{theorem} \label{thm1}
  Under Assumption~\ref{main.Condition}, the sequence of rescaled volatility processes $\{V^{(n)} \}_{n \geq 1}$ is C-tight. Moreover, any accumulation point  $\{V_*(t);t\geq 0\}$ in $\mathbf{D}([0,\infty);\mathbb{R}_+)$ satisfies the following stochastic Volterra equation:
  \beqlb\label{eq1}
   V_*(t)= V_*(0) \cdot \big(1-F^{\alpha,\gamma}(t)\big)+ \int_0^t \frac{a}{b}\cdot f^{\alpha,\gamma}(s)ds+ \int_0^t  f^{\alpha,\gamma}(t-s)\cdot \frac{1}{b}\sqrt{V_*(s)}dB(s),
   \quad  t\geq 0, 
  \eeqlb
 where $B$ is a standard Brownian motion. 
  The solution to the above equation is unique in law. 
  In particular, the sequence of rescaled volatility processes converges in law. 
 \end{theorem}
 
 Existence and uniqueness in law of the stochastic Volterra equation \eqref{eq1} has been established in \cite{JaberLarssonPulido2019}. Our contribution is the proof of convergence of the rescaled volatility process. Any solution to this equation can be equivalently rewritten in a more familiar form. The proof can be carried out as in \cite{ElEuchRosenbaum2019b}.  We provide an alternative and simpler proof in Section 3 by solving the corresponding Wiener-Hopf equation.

 \begin{theorem} \label{thmfractional} 
 	Any solution to the stochastic Volterra equation \eqref{eq1} can be equivalently represented as
  \beqlb \label{eqn.FracVol}
  V_*(t) = V_*(0)+  \int_0^t\frac{(t-s)^{\alpha-1}}{\Gamma(\alpha)} \cdot \gamma \Big(\frac{a}{b}-V_*(s)\Big)ds+\int_0^t \frac{(t-s)^{\alpha-1}}{\Gamma(\alpha) }\cdot \frac{\gamma}{b}\sqrt{V_*(s)}dB(s), \quad t \geq 0.
  \eeqlb 
 \end{theorem}

 \subsubsection{The price process}
 
 The preceding theorem shows that the intensity of our order arrivals process converges in law to the unique weak solution of a stochastic Volterra equation. In this section we show that the convergence of the intensity process implies the weak convergence of the sequence of rescaled price processes defined by 
 \beqnn
 P^{(n)}(t):= \frac{P_{n}(nt)}{n^\alpha}, \quad t \geq 0. 
 \eeqnn
 
 In terms of the compensated Poisson random measure $\widetilde{N}_{n,0}(ds, du, dz) := N_{n,0}(ds, du, dz)-ds\nu_n(du)dz$, the rescaled price process $P^{(n)}$ satisfies the stochastic integral equation
 
 \begin{equation}\label{rescaledP}
 \begin{split}
 	P^{(n)}(t) & = \frac{P_{n,0}}{n^\alpha}+  n^{\alpha}\int_{\mathbb R} u\, \nu_n(du) \cdot \int_0^tV^{(n)}(s)ds \\ & ~~~ + \int_0^{t} \int_{\mathbb R} \int_0^{V^{(n)}(s-)}  \frac{u}{n^\alpha} \widetilde N_{n,0}(n\cdot ds, du, n^{2\alpha-1}\cdot dz), \quad t \geq 0. 
 \end{split}
 \end{equation}
 To obtain a non-degenerate limit for the sequence $\{ P^{(n)} \}_{n\geq 1}$, it is natural to assume that the sequence of probability measures $\{\nu_n\}_{ n \geq 1}$ satisfies the following condition. 
 
 \begin{assumption}\label{main.Condition.02}
 There exist four constants $P_*(0),\sigma_p\geq 0$, $b_p\in\mathbb{R}$ and $\epsilon > 0$ such that
 	\beqnn
 	\frac{P_{n,0}}{n^\alpha} \to P_*(0),\quad 
 	n^\alpha \int_\mathbb{R} u\, \nu_n(du) \to b_p,\quad 
 	 \int_\mathbb{R} |u|^2 \nu_n(du) \to \sigma^2_p, \quad
	 \sup_{n\geq 1}\int_{\mathbb R} |u|^{2+\epsilon} \nu_n(du)<\infty.
 	\eeqnn
 	\end{assumption}
 
 We are now ready to state our convergence result for the price process. The proof is given in Section \ref{Sec.proofs} below. 
 
 \begin{theorem}\label{MainThm.03}
  Let $V_*$ be any solution to the stochastic integral equation \eqref{eq1}. 
  Under Assumption~\ref{main.Condition.02}, the sequence of rescaled price processes $\{P^{(n)}\}_{ n \geq 1}$ converges in law to the unique solution of the stochastic differential equation 
 	\beqlb\label{SL.Price}
 	P_*(t) = P_*(0) + \int_0^t b_p V_*(s)ds + \int_0^t \sigma_p\sqrt{V_*(s)}dW(s), \quad t \geq 0,
 	\eeqlb
 	where the standard Brownian motion  $W$  is independent of $B$. 
 	In particular, under Assumption~\ref{main.Condition} and \ref{main.Condition.02} the sequence $\{(P^{(n)} , V^{(n)} ) \}_{n \geq 1}$ converges in law to the unique solution to  the coupled system of stochastic equations (\ref{eq1}) and (\ref{SL.Price}). 
  
 \end{theorem}
  
 The preceding theorem shows that the sequence of rescaled price-volatility processes converges in law to a rough Heston model driven by two {\sl independent} Brownian motions. In the light-tailed cases considered in \cite{HorstXu2022, JaissonRosenbaum2015}  and in the heavy-tailed case/rough model considered in \cite{ElEuchFukasawaRosenbaum2018, GatheralKeller2019} the Brownian motions driving the price and volatility process are correlated. The reason for the independence of the Brownian motions in our model is two-fold. First, in the heavy-tailed case the price and volatility processes are scaled differently in space whereas the scaling is the same in light-tailed case considered in \cite{JaissonRosenbaum2015}. Second, unlike in \cite{HorstXu2022} in our model price changes are conditionally deterministic.

 \begin{remark}
 Correlation between the driving Brownian motions can be introduced into our model by correlating price and volatility changes as in  \cite{HorstXu2022} and to consider a dynamics of the form 
 \beqnn
 P_t\ar=\ar P_0+\int_0^t \int_{\mathbb{R}^2}  u_1N(ds,du), \cr
 V_t\ar=\ar \mu_t + \int_0^t\int_{\mathbb{R}^2} u_2  \cdot \phi(t-s) N(ds,du), 
 \eeqnn
 where 
 \[
 	N(dt,du) := \sum_{k=1}^\infty \mathbf{1}_{\{ \tau_k \in dt, \xi_k\in du  \}}
 \]
is now a Hawkes random measure on $[0,\infty)\times \mathbb{R}^2$ with intensity $V_{t-}dt\nu(du)$. Here $\{\tau_k\}$ describes again the arrivals of market orders and the ${\mathbb R}^2$-valued sequence $\{\xi_k\}$ describes the joint changes in prices and volatilities caused by the respective order arrival. Introducing this more general dynamics results in a more cumbersome model analysis as we would have to consider the joint dynamics of prices and volatilities right from the start instead of analyzing the processes separately as we do in what follows. At the same time, introducing correlated movements would not result in any new mathematical challenges. As our focus is on the convergence of the volatility process we prefer to work within the more convenient setting of independent driving noises.
\end{remark}


%% file: ConvergenceVol.tex
 \section{Proofs}\label{Sec3}
 \setcounter{equation}{0}
 
 In this section we prove the weak convergence of the rescaled models to a fractional Heston stochastic volatility model. The main challenge is to prove the $C$-tightness of the sequence of rescaled volatility processes.  
%
 The rescaled volatility processes can be expressed as
 \beqlb\label{decomV}
 	V^{(n)}(t) = V^{(n)}_0(t)  + I^{(n)}(t) + J^{(n)}(t), \quad t \geq 0,  
 \eeqlb
 where the processes on the right-hand side of the above equation are defined by
 \begin{equation}
 \begin{split}
 V^{(n)}_0(t) & := \frac{1}{n^{2\alpha-1}} \big( \Lambda_n(nt) + R_n*\Lambda_n (nt)\big), \\ 
 I^{(n)}(t) & :=  \frac{\mu_n}{n^{2 \alpha - 1}} + \frac{\mu_n}{n^{2\alpha-1}} \cdot \int_0^{nt}R_n(s)ds  , \\
 J^{(n)}(t) & := \int_0^t  \int_0^{V^{(n)}(s-)}  \frac{R_n\big(n(t-s)\big) }{n^{2\alpha-1}}   \widetilde{N}^{(n)}_{1}(ds,dz),  \label{eqn.Jn}
 \end{split}
 \end{equation}
 and where 
 \beqnn
 \widetilde{N}^{(n)}_{1}(ds,dz):=\widetilde{N}_{n,1}(n\cdot ds ,n^{2\alpha-1}\cdot dz)
 \eeqnn 
 is a compensated Poisson random measure on $\mathbb{R}_+^2$ with intensity $n^{2\alpha}\cdot dsdz$. 
 
 The $C$-tightness of the first two processes  is obtained in Section \ref{Sec:V0}. To obtain the tightness of the rescaled volatility processes, by Corollary~3.33(1) in \cite[p.353]{JacodShiryaev2003} it then remains to establish the $C$-tightness of the stochastic integral processes, which is more challenging.  Two steps are required to overcome the challenge. 
 \begin{enumerate}
 	\item[$\bullet$] In Section~\ref{Subsection.C-tight}, we introduce a novel technique to verify the $C$-tightness
	of {\sl c\`{a}dl\`{a}g} processes based on the classical Kolmogorov-Chentsov tightness criterion for {\sl continuous} processes. 
 	
 	\item[$\bullet$] In Section~\ref{Subsection.Estimates}, we establish priori estimates, including uniform estimates for the resolvents $\{ R_n \}_{n\geq 1}$ and their derivatives and a uniform moment estimate of all orders for $\{ J^{(n)} \}_{n\geq 1}$ and $\{ V^{(n)} \}_{n\geq 1}$.
 \end{enumerate}

We then proceed to establish the tightness of the sequence $\{ J^{(n)} \}_{n\geq 1}$ in Section \ref{Sec.tight} and its weak convergence in Section \ref{Sec.conv}. Our main theorems are proved in Section \ref{Sec.proofs}. Section \ref{Sec.counter} contains a counterexample that shows that the weak convergence of the integrated volatility processes do in general not imply the weak convergence of the volatility processes themselves. 


 \subsection{$C$-tightness of $\{V_0^{(n)} \}_{n\geq 1}$ and $\{I^{(n)} \}_{n\geq 1}$}\label{Sec:V0}
  
 As a preparation, in the next proposition we provide an asymptotic result for the rescaled resolvent sequence $\{n^{1-\alpha} R_n(n\cdot )\}_{n\geq 1}$ and their integral functions.

 \begin{proposition} \label{Prop.ConvergenceRn}
 Recall the constant $b$ defined in Assumption~\ref{main.Condition}(3). 
 As $n\to\infty$, we have
 \beqnn
 \sup_{t\geq 0} \Big|\int_0^t n^{1-\alpha} R_n(ns)ds-   \frac{F^{\alpha,\gamma}(t)}{b} \Big| \to 0. 
 \eeqnn 
 \end{proposition}
 \proof Similarly as in the proofs of Lemma 4.3 \cite{JaissonRosenbaum2016} and Lemma 4.4 in \cite{Xu2021b}, the main idea is to show that the Laplace transform of the function $R^{(n)}$ converges to the Laplace transform of the Mittag-Leffler density function $ b^{-1}\cdot f^{\alpha,\gamma}$.
 
 The resolvent $R_n$ satisfies resolvent equation
 \beqlb\label{eqn.Rn}
 R_n(t)= \zeta_n \cdot \phi(t)+ \zeta_n \cdot \phi * R_n(t),\quad t\geq 0.
 \eeqlb
 Taking Laplace transform on both sides of this equation shows that 
 \beqnn
 \mathcal{L}_{R_n}(\lambda) := \int_0^\infty e^{-\lambda t} R_n(t)dt = \frac{\zeta_n\cdot \mathcal{L}_{\phi}(\lambda)}{1-\zeta_n\cdot \mathcal{L}_{\phi}(\lambda)},
 \quad \lambda \geq 0. 
 \eeqnn
 By a change of variables,  
 \beqlb\label{eqn.LaplaceR}
 \int_0^\infty e^{-\lambda t}n^{1-\alpha} R_n(nt)dt
 = \frac{\mathcal{L}_{R_n}(\lambda/n) }{n^\alpha}  
 = \frac{\zeta_n\cdot \mathcal{L}_{\phi}(\lambda/n) }{n^\alpha(1-\zeta_n)
 	+ \zeta_n \cdot n^\alpha \big (1-  \mathcal{L}_{\phi}(\lambda/n) \big )},\quad \lambda \geq 0. 
 \eeqlb
 The monotonicity of $\mathcal{L}_\phi$ and the assumption that $\zeta_n\to 1$ induce that  
 \beqnn
 \lim_{n\to\infty}  \zeta_n\cdot \mathcal{L}_{\phi}(\lambda/n) = 1,
 \quad \lambda >0. 
 \eeqnn
 To analyze the asymptotics of the last denominator in (\ref{eqn.LaplaceR}), we use the fact that the tail-distribution $\overline\varPhi$ defined in (\ref{eqn.TailvarPhi}) is regularly varying with index $-\alpha$. 
 An application of Corollary 8.1.7 in \cite[p.334]{BinghamGoldieTeugels1987}  shows that 
 \beqnn
  1- \mathcal{L}_{\phi}(z) 
  \sim \Gamma(1-\alpha)  \overline\varPhi(1/z),
  \quad \mbox{as }z\to 0+.
 \eeqnn
 By \eqref{eqn.TailvarPhi}, we have $\overline\varPhi(n/\lambda )= (1+\sigma\cdot n/\lambda )^{-\alpha} \sim n^{-\alpha}\cdot \lambda^{\alpha}/\sigma^\alpha $ as $n\to\infty$ and then conclude that 
 \beqnn
  \lim_{n\to\infty} n^{\alpha}\big (1-\mathcal{L}_{\phi}(\lambda/n)\big ) = \frac{\Gamma(1-\alpha)}{\sigma^\alpha}\cdot\lambda^\alpha,\quad \lambda >0. 
 \eeqnn
 Moreover, by Assumption~\ref{main.Condition}(3), $n^\alpha(1-\zeta_n) \to b$. Plugging this into the denominator in (\ref{eqn.LaplaceR}) yields that
 \beqnn
 \lim_{n\to\infty} \int_0^\infty e^{-\lambda t}n^{1-\alpha} R_n(nt)dt
 = \frac{\sigma^\alpha}{b\sigma^\alpha+\Gamma(1-\alpha)\lambda^{\alpha}} 
 = \mathcal{L}_{b^{-1}\cdot f^{\alpha,\gamma}}(\lambda),
  \quad \lambda >0.
 \eeqnn
 
 The pointwise convergence of the Laplace transform of $ n^{1-\alpha} R_n(n\cdot)$ to that of $b^{-1} \cdot f^{\alpha,\gamma}$ yields the weak convergence of the finite measure with density function $n^{1-\alpha} R_n(n\cdot)$ to the finite measure with density function $b^{-1} \cdot f^{\alpha,\gamma}$, which, together with the continuity of $F^{\alpha,\gamma}$,  induces the desired uniform convergence of the integral function of $n^{1-\alpha} R_n(n\cdot)$ to $b^{-1} \cdot F^{\alpha,\gamma}$. 
  \qed
  
 
 
 \begin{corollary}\label{Cor.ConvergeVI}
 As $n\to\infty$, we have 
  \beqnn
  \sup_{t\geq 0}\Big|  V^{(n)}_0(t) -V_*(0) \big( 1-F^{\alpha,\gamma}(t) \big)  \Big| \to 0
  \quad \mbox{and}\quad
  \sup_{t\geq 0} \Big| I^{(n)}(t) - \frac{a}{b} \cdot F^{\alpha,\gamma}(t) \Big| \to 0. 
  \eeqnn 
 \end{corollary}
 \proof The second uniform convergence result is a direct consequence of Proposition~\ref{Prop.ConvergenceRn} and Condition~\ref{main.Condition}(3).
 For the first one, by Assumption~\ref{main.Condition}(2) we have uniformly in $t\geq 0$,
 \beqnn
 V^{(n)}_0(t)\ar =\ar \frac{V_{n,0}}{n^{2\alpha-1}} \big(  \overline\varPhi(nt) + R_n*\overline\varPhi(nt)\big)
 \sim V_*(0)  \cdot \big(  \overline\varPhi(nt) + R_n*\overline\varPhi(nt)\big),
 \eeqnn
 as $n\to\infty$.
 It remains to prove that $\overline\varPhi(nt) + R_n*\overline\varPhi(nt)$ converges uniformly to $ 1-F^{\alpha,\gamma}(t)$. 
 Integrating both sides of (\ref{eqn.Rn}) on $[t,\infty)$ and then dividing them by $\zeta_n$,
 \beqlb\label{eqn.3001}
 \frac{1}{\zeta_n}\int_t^\infty R_n (s) ds
 \ar=\ar \overline\varPhi(t) + \int_t^\infty \phi *R_n(s)ds
 = \overline\varPhi(t) + \int_0^\infty \phi *R_n(s)ds- \int_0^t  \phi *R_n(s)ds.
 \eeqlb
 By using Fubini's lemma together with the fact that $\|\phi\|_{L^1}=1$, we have
 \beqnn
  \int_0^\infty \phi *R_n(s)ds
  = \int_0^\infty R_n(s) ds \int_0^\infty  \phi(r)dr 
  =\int_0^\infty R_n(s) ds 
 \eeqnn
 and 
 \beqnn
\int_0^t  \phi *R_n(s)ds
 \ar=\ar \int_0^t R_n(t-s) \int_0^{s}  \phi(r)dr ds= \int_0^t R_n(t-s) \big(1- \overline\varPhi(s) \big) ds
 = \int_0^t R_n(s)ds -   \overline\varPhi* R_n(t).
 \eeqnn
  Taking these two results back into (\ref{eqn.3001}), we see that
 \beqnn
 \frac{1}{\zeta_n}\int_t^\infty R_n (s) ds
 \ar=\ar 
 \overline\varPhi(t) + \int_t^\infty R_n(s)ds + \overline\varPhi* R_n(t). 
 \eeqnn
 Moving the second term on the right side to the left side,
 \beqnn
 \overline\varPhi(t) +  \overline\varPhi* R_n(t) = \frac{1-\zeta_n}{\zeta_n} \int_t^\infty R_n (s) ds . 
 \eeqnn
 By a change of variables, Proposition~\ref{Prop.ConvergenceRn} and Assumption~\ref{main.Condition}(3), 
 \beqnn
  \overline\varPhi(nt) +  \overline\varPhi* R_n(nt) = \frac{n^{\alpha}(1-\zeta_n)}{\zeta_n} \int_t^\infty n^{1-\alpha} R_n (ns) ds 
  = \frac{n^{\alpha}(1-\zeta_n)}{\zeta_n} \cdot \Big(\mathcal{I}_{R^{(n)}} (\infty)  - \mathcal{I}_{R^{(n)}} (t)  \Big) , 
 \eeqnn
  which converges uniformly to $1-F^{\alpha,\gamma}(t)$ as $n\to\infty$. 
  \qed

 \subsection{A $C$-tightness criterion for c\`{a}dl\`{a}g processes}
 \label{Subsection.C-tight}
 
 In this section we provide a novel $C$-tightness criterion for a sequence of stochastic processes $\{X^{(n)}\}_{n\geq 1}$ in $\mathbf{D}(\mathbb{R}_+;\mathbb{R})$ by using the well-known Kolmogorov-Chentsov tightness criterion for continuous processes; see Problem~4.11 in \cite[p.64]{KaratzasShreve1988}. 
 
 The Kolmogorov-Chentsov condition states that a sequence of continuous processes $\{X^{(n)}\}_{n\geq 1}$ is tight if for each $T\geq 0$, there exist constants $C,\beta,p>0 $ and $\kappa>1$ such that for any $0\leq t_1<t_2\leq T$, 
 \beqlb\label{eqn.KolTight}
 \sup_{n\geq 1}\mathbf{E}\Big[\big|X^{(n)}(0)\big|^{\beta}\Big]<\infty
 \quad\mbox{and}\quad  
 \sup_{n\geq 1} \mathbf{E}\Big[\big|X^{(n)}(t_1)-X^{(n)}(t_2)\big|^{p}\Big]\leq C\cdot \big|t_2-t_1\big|^{\kappa} . 
 \eeqlb
  
In what follows we prove that a sequence of c\`adla\`g processes is $C$-tight if it can be approximated by a sequence of $C$-tight processes. Our notion of approximation is asymptotic indistinguishability as introduced in the following definition. 
   
 \begin{definition}
 We say a sequence of stochastic processes $\{Y^{(n)} \}_{n\geq 1}$ in $\mathbf{D}( \mathbb{R}_+; \mathbb{R})$ is {\rm asymptotically indistinguishable} from   $\{X^{(n)} \}_{n\geq 1}$
 if for each $T\geq 0$, as $n\to\infty$,
 \beqlb\label{eqn.Indistinguishable}
 \sup_{t\in[0,T]}|X^{(n)}(t)-Y^{(n)}(t)|  \overset{\rm p}\to 0 .
 \eeqlb
 \end{definition}
 
 As a direct consequence of Theorem~3.1 in \cite[p.27]{Billingsley1999}, the next proposition shows that tightness and $C$-tightness are properties shared by asymptotically indistinguishable sequences of stochastic processes.

 \begin{proposition} \label{Prop.TightnessEquiv}
  The sequence $\{X^{(n)} \}_{n\geq 1}$ is tight [resp. $C$-tight] if and only if one and hence every asymptotically indistinguishable process sequence $\{Y^{(n)} \}_{n\geq 1}$ is tight [resp. $C$-tight].
 \end{proposition}
 
For a given sequence of stochastic processes  $\{X_n\}_{n\geq 1}$  we now construct piece-wise linear approximations that allow us to verify the tightness of  $\{X_n\}_{n\geq 1}$ under suitable moment conditions on its jumps, akin to the moment conditions in the Kolmogorov-Chentsov tightness criterion for {\sl continuous} processes. 

In what follows we denote by $\Delta_h$ the forward difference operator with step size $h>0$, i.e.,
 \beqnn
 \Delta_hf(x) := f(x+h) -f(x),\quad x\in\mathbb{R}. 
 \eeqnn 
 Moreover, we fix some $\theta > 2$ and define the following piece-wise linear approximations on the grids $\{ \frac{k}{n^\theta} : k=0,1,\cdots\}$: 
 \beqlb\label{interpolation}
 X^{(n)}_\theta(t):= X^{(n)}\big( [n^\theta t]/n^\theta\big)+   \Delta_{1/n^\theta} X^{(n)}\big( [n^\theta t]/n^\theta\big) \cdot \big(n^\theta t-[n^\theta t] \big), 
 \quad t\geq 0.
 \eeqlb
 
 {The next lemma is key to our analysis. It provides sufficient conditions on the sequence  $\{X^{(n)} \}_{n\geq 1}$ that guarantee that the above linear interpolation forms as a tight approximation. }
 
 \begin{lemma} \label{tightness condition}
If $$\sup_{n \geq 1}  \mathbf{E} \big[|X^{(n)}(0)|^\beta\big] < \infty$$ for some $\beta>0$, then
the sequence $\{X^{(n)} \}_{n\geq 1}$  is $C$-tight if 
 for any $T\geq 0$ and some constant $\theta>2$, the following two  conditions hold.
 \begin{enumerate}
 	\item[(1)] $\displaystyle\sup_{k=0,1,\cdots,[Tn^\theta]} \sup_{h\in[0,1/n^{\theta}]} \big|\Delta_h X^{(n)}(k/n^\theta) \big|  \overset{\rm p}\to 0$  as $n\to\infty$.
 	
 	\item[(2)] There exist some constants $C >0 $, $p\geq 1$, $m\in\{1,2, ...\}$ and pairs $\{ (a_i,b_i) \}_{i=1,\cdots ,m}$ satisfying 
 	\beqnn
 	a_i\geq 0,\quad b_i>0,\quad \rho:=\min_{1\leq i\leq m} \big\{ b_i + a_i/\theta \big\}>1 , 
 	\eeqnn
    such that for all $n\geq 1$ and $h \in (0,1)$,
 	\beqlb \label{condition1}
 	 \sup_{t\in [0,T]}	\mathbf{E}\Big[\big|\Delta_h X^{(n)}(t)\big|^{p}\Big]\leq 
 		C\cdot \sum_{i=1}^m \frac{h^{b_i}}{n^{a_i}}. 
 	\eeqlb 
 \end{enumerate} 

 \end{lemma}
 \proof
 We first prove that condition $(1)$ guarantees that the sequence of linear interpolations $\{ X^{(n)}_\theta\}_{n\geq 1}$ defined in \eqref{interpolation} is asymptotically indistinguishable from $\{ X^{(n)}\}_{n\geq 1}$. 
 In fact, for any $t\in[0,T]$ we have that
 \beqnn
 X^{(n)}(t)-X^{(n)}_\theta(t) = \big(X^{(n)}(t)-X^{(n)}([n^\theta t]/n^\theta)\big) -
 \Delta_{1/n^\theta} X^{(n)}_\theta([n^\theta t]/n^\theta) \cdot \big(n^\theta t-[n^\theta t] \big) 
 \eeqnn
 and then by the triangle inequality,
 \beqnn
 \big| X^{(n)}(t)-X^{(n)}_\theta(t)  \big| \leq 2\cdot \sup_{h\in[0,1/n^{\theta}]} \big|\Delta_h X^{(n)}([n^\theta t]/n^\theta) \big|.
 \eeqnn
 This shows that 
 \beqnn
 \sup_{t\in[0,T]} \big| X^{(n)}(t)-X^{(n)}_\theta(t)  \big| \leq 2\cdot \sup_{k=1,\cdots, [n^\theta t]} \sup_{h\in[0,1/n^{\theta}]} \big|\Delta_h X^{(n)}(k/n^\theta) \big|,
 \eeqnn
 which goes to $0$ in probability as $n\to\infty$ because of condition $(1)$.   
 
 By Proposition~\ref{Prop.TightnessEquiv}, it remains to verify that the sequence $\{ X^{(n)}_\theta \}_{n\geq 1}$ satisfies the classical Kolmogorov-Chentsov tightness criterion {for continuous processes}. The first moment condition in (\ref{eqn.KolTight}) holds because  \beqnn
 \sup_{n \geq 1}  \mathbf{E} \Big[\big|X^{(n)}_\theta(0)\big|^\beta\Big]=\sup_{n \geq 1}  \mathbf{E} \Big[\big|X^{(n)}(0)\big|^\beta\Big] < \infty.
 \eeqnn 
 To verify that condition (2) implies the second moment condition in (\ref{eqn.KolTight}), it is enough to prove that for some constants $C>0$ and $\kappa \in (1, \rho)$ such that uniformly in $h\in (0,1)$ and $t\in[0,T]$,
 \beqnn
  \sup_{n\geq 1}  \mathbf{E}\Big[\big|\Delta_h X^{(n)}_\theta(t)\big|^{p} \Big]
 \leq C\cdot h^{\kappa}. 
 \eeqnn
 To this end, we fix $t \in [0,T]$ and $h \in(0,1)$ and distinguish the following two cases. 
 
 \medskip
 
 {\bf Case 1.} If $h\leq n^{-\theta}$, there exists an integer $k\geq 0$ such that $t\in[\frac{k}{n^\theta},\frac{k+1}{n^\theta}]$ and hence $t+h\in[\frac{k}{n^\theta},\frac{k+2}{n^\theta}]$. Thus,
 \beqnn
 \big|\Delta_hX^{(n)}_\theta(t) \big|\leq hn^\theta\cdot \Big( 
 \big|\Delta_{1/n^\theta} X^{(n)}  (k/n^\theta)\big| +
 \big|\Delta_{1/n^\theta}X^{(n)} \big((k+1)/n^\theta \big)\big|\Big).
 \eeqnn
 Raising both sides of this inequality to the $p$ power, then using the power mean inequality and finally taking expectations, we have 
 \beqnn
 \mathbf{E}\Big[\big|\Delta_h X^{(n)}_\theta(t) \big|^p\Big] 
 \ar\leq\ar 2^p\cdot  (hn^\theta)^p\cdot \Big( 
 \mathbf{E}\Big[\big|\Delta_{1/n^\theta}X^{(n)} (k/n^\theta)\big|^p\Big] +
 \mathbf{E}\Big[\big|\Delta_{1/n^\theta}X^{(n)} \big((k+1)/n^\theta \big)\big|^p\Big]\Big) . 
 \eeqnn
 By condition \eqref{condition1} and then the two facts that $h\leq n^{-\theta}$ and $\theta \kappa-a_i-\theta b_i \leq \theta(\kappa-\rho)<0 $, there exists a constant $C > 0$ that is independent of $t,h$ and $n$ such that 
 \beqnn
 \mathbf{E}\Big[\big|\Delta_h X^{(n)}_\theta(t) \big|^p\Big] 
  \leq   C  \sum_{i=1}^m n^{\theta p-a_i-\theta b_i} \cdot h^{p-\kappa} \cdot h^{\kappa} 
  \leq  C  \sum_{i=1}^m n^{\theta \kappa-a_i-\theta b_i } \cdot h^{\kappa} 
  \leq  C \cdot h^\kappa.
 \eeqnn

 \medskip
 
 {\bf Case 2.} If $h>n^{-\theta}$, there exist two integers $k< l$ such that $t\in [\frac{k}{n^\theta},\frac{k+1}{n^\theta} ]$ and $t+h\in[\frac{l}{n^\theta},\frac{l+1}{n^\theta}]$. 
 By the triangle inequality, we can bound $ \big|\Delta_hX^{(n)}_\theta(t) \big|$ by
 \beqnn
  \big| X^{(n)}_\theta(t+h) -X^{(n)}_\theta(l/n^\theta) \big| + \big| X^{(n)}_\theta((k+1)/n^\theta) -X^{(n)}_\theta(t) \big| + \big|X^{(n)}_\theta(l/n^\theta) - X^{(n)}_\theta((k+1)/n^\theta) \big| . 
 	\eeqnn
 Applying our result in {\bf Case 1} to  the first two terms, there exists a constant $C > 0$ that is independent of $t,h$ and $n$ such that
 \beqnn
 \mathbf{E}\Big[  \big| X^{(n)}_\theta(t+h) -X^{(n)}_\theta(l/n^\theta) \big|^p\Big] + \mathbf{E}\Big[\big| X^{(n)}_\theta((k+1)/n^\theta) -X^{(n)}_\theta(t) \big|^p \Big] \leq C\cdot h^\kappa.
 \eeqnn
 For the third term, by definition of $X^{(n)}_\theta$ we have 
 \beqnn
 \big|X^{(n)}_\theta(l/n^\theta) - X^{(n)}_\theta((k+1)/n^\theta) \big| 
 \ar=\ar \big| X^{(n)}(l/n^\theta) - X^{(n)}((k+1)/n^\theta) \big| . 
 \eeqnn
 Applying first the moment estimate \eqref{condition1} to the last term, then the inequality $h > 1/n^\theta$ and finally the fact that $1<\kappa<\rho$, we again obtain that for some constant $C>0$ that is independent of $t,h$ and $n$ such that
 \beqnn
 \mathbf{E}\Big[ \big| X^{(n)}_\theta(l/n^\theta) - X^{(n)}_\theta((k+1)/n^\theta)  \big|^p \Big] 
 \leq 	C\cdot \sum_{i=1}^m n^{-a_i}\Big(\frac{l-k-1}{n^\theta}\Big)^{b_i}  \leq 	C\cdot \sum_{i=1}^m h^{b_i + a_i/\theta } \leq C \cdot h^{\kappa}.
 \eeqnn 
 which, together with the preceding estimates, induces that 
 $ \mathbf{E}\big[\big|\Delta_h X^{(n)}_\theta(t) \big|^p\big] 
 \leq   C \cdot h^\kappa$. 
 \qed
 
 
 \subsection{A priori estimates}
 \label{Subsection.Estimates}
 
 In this section we provide priori estimates that will be important to our subsequent analysis. We start with a uniform upper bound on the resolvent sequence $\{R_n\}_{n\geq 1}$. 
 
 \begin{proposition}\label{boundR}
 There exists a constant $C>0$ such that for any $t\geq 0$,
 \beqlb \label{ubRH}
 	\sup_{n\geq 1} R_n(t) \leq C\cdot(1+t)^{\alpha-1}. 
 \eeqlb
 \end{proposition}
 \proof Let $R_\alpha$ be the resolvent associated with the kernel $\phi$, which is defined as the  unique solution of (\ref{eqn.Resolvent}) with $\zeta=1$. 
 The two functions  $R_n$ and $R_\alpha$ admit the respective Neumann series expansions
 \beqlb\label{eqn.NeumannSeries}
 R_n= \sum_{k=1}^\infty (\zeta_n\cdot \phi)^{*k} 
 \quad \mbox{and} \quad
 R_\alpha= \sum_{k=1}^\infty   \phi^{*k}  . 
 \eeqlb
 The convergence of the first series follows from the fact that $\zeta_n \|\phi\|_{L^1}= \zeta_n<1$. For the second series, let $\phi_\beta(t) := e^{-\beta t}\cdot \phi (t)$ for $t\geq 0$ and $\beta>0$. It is easily verified that $(\phi_\beta)^{*k}(t) = e^{-\beta t}\cdot \phi^{*k} (t)$ for $t\geq 0$. Since $\|\phi_\beta\|_{L^1}<1$ and $\|(\phi_\beta)^{*k}\|_{L^1}=\|\phi_\beta\|_{L^1}^k$, we have as $m\to\infty$ that
 	\beqnn
 	\sum_{k=1}^m   \phi^{*k}(t) = e^{\beta t}\sum_{k=1}^m e^{-\beta t}  \phi^{*k}(t) 
 	\ar=\ar e^{\beta t}\sum_{k=1}^m  (\phi_\beta)^{*k}(t) \cr
 	\ar\to\ar e^{\beta t}\sum_{k=1}^\infty  (\phi_\beta)^{*k}(t) = \sum_{k=1}^\infty e^{\beta t} \cdot e^{-\beta t}\cdot \phi^{*k} (t) 
 	=\sum_{k=1}^\infty \phi^{*k} (t) ,\quad t\geq 0. 
 	\eeqnn
 The stability condition $\zeta_n<1$ induces that $R_n< R_\alpha$. 
 It is obvious that $	R_\alpha$ is continuous with $	R_\alpha(0)=\alpha \sigma <\infty$.  
 Applying Karamata's theorem (see Theorem 8.1.6 in \cite[p.333]{BinghamGoldieTeugels1987}) to the Laplace transform of $R_\alpha$  shows that as $ t\to\infty$, 
 \beqlb\label{eqn.0014}
 	R_\alpha(t)\sim C \cdot t^{\alpha-1}  
 \eeqlb
 for some constant $C>0$. 
 Hence, \eqref{ubRH} holds; see Proposition 4.8 in \cite{Xu2021b} for more details.
 \qed 
 
 Additionally, in the fore-coming tightness proofs we shall also need a uniform upper bound on the derivative sequence $\{R'_n\}_{n\geq 1}$ of the resolvents $\{R_n\}_{n\geq 1}$; see the next proposition. 
  
 \begin{proposition} \label{propG}
 There exists a constant $C>0$ such that uniformly in $t\geq 0$,
 \beqlb \label{ubGH}
  \sup_{n\geq 1}	\big|R'_n(t) \big|\leq C \cdot (1+t)^{\alpha-2}. 
 \eeqlb
 \end{proposition}
 \proof  The kernel $\phi$ introduced in \eqref{eqn.Kernel} is completely monotone, that is,  
 \beqnn
 (-1)^k\frac{d^k}{dt^k}\phi(t) \geq 0,\quad k\in \{0, 1,2 ... \}, \ t\geq 0,
 \eeqnn
 it follows from the well-known Bernstein's theorem on completely monotone functions that it can be uniquely written as the Laplace transform of some non-negative measure $\mu_1$, that is,
 \beqnn
  \phi(t)=\int_0^\infty e^{- xt}\mu_1(dx), \quad t \geq 0. 
 \eeqnn	
 Similarly, there exists a unique non-negative measure $\mu_k$ such that the $k$-th order convolution of $\phi$ can be written as 
 \beqnn
  \phi^{*k}(t)=\int_0^\infty e^{- xt}\mu_k(dx), \quad t \geq 0.
 \eeqnn
 Plugging these back into the two Neumann series expansions in (\ref{eqn.NeumannSeries}),
 \beqlb\label{eqn.0013}
 R_n(t)
   =\sum_{k=1}^\infty \int_0^\infty  |\zeta_n|^k e^{-xt}\mu_k(dx)
 \quad \mbox{and}\quad 
 R_\alpha(t) =\sum_{k=1}^\infty \int_0^\infty    e^{-xt}\mu_k(dx),	 	
 \quad t\geq 0.
 \eeqlb
 Since $0<\zeta_n<1$, these allow us to bound our derivatives $R'_n$ by the derivative $R'_\alpha$ of the resolvent $R_\alpha$, i.e., for any $n \geq 1$ and $t \geq 0$,
 \beqnn
  \big| R'_n(t) \big|
  = \int_0^\infty x\cdot \sum_{k=1}^\infty |\zeta_n |^k  e^{-xt} \mu_k(dx)
  \leq\int_0^\infty x\cdot e^{-xt}\sum_{k=1}^\infty \mu_k(dx)
  = \big|R'_\alpha(t)\big|. 
 \eeqnn
 From (\ref{eqn.NeumannSeries}), we have $R'_\alpha= \phi'+ \phi' *\sum_{k=1}^\infty \phi^{*k}$ and hence $|R'_\alpha(0)|= |\phi'(0)|=\alpha \sigma  (\alpha+1)<\infty$. 
 Finally, an application of Bernstein's theorem and (\ref{eqn.0013}) shows that $R_\alpha$ is completely monotone and hence $R'_\alpha$ is monotone.  
 By the regular variation of  $R_\alpha$ at infinity (see (\ref{eqn.0014})) and Proposition 2.5(b) in \cite{Resnick2007} we can obtain that as $t\to\infty$,
 \beqnn
  \big|R'_\alpha(t) \big|\sim C\cdot t^{\alpha-2} 
 \eeqnn
 for some constant $C>0$
 and hence the desired uniform upper bound follows immediately. 
  \qed

 The next proposition provides moment estimates for stochastic integrals driven by Poisson random measures when one of the upper integral boundaries is described by stochastic processes. 
 It is a direct corollary of Lemma D.1 in \cite{Xu2021b} and the proof is omitted.
 
 \begin{proposition} \label{2pine}
 For $p\geq 1$ and $T\geq 0$, assume that $f$ is a measurable function in $L^2([0,T];\mathbb{R})$, and  $X:=\{X(t):t\geq 0\}$ is a non-negative $(\mathscr{F}_t)$-c\`adl\`ag process  with
 \beqnn
 L:= \sup_{t\in[0,T]}\textbf{E}\Big[\big|X(t)\big|^p\Big] <\infty. 
 \eeqnn
 Let $\widetilde{N}_\eta(ds,dz)$ be a compensated Poisson random measure on $\mathbb{R}_+^2$ with intensity $\eta \cdot dsdz$ for some constant $\eta>0$. 
 Then there exists a constant $C>0$ that depends only on $p$, $T$ and $L$ such that 
 \begin{equation}
 	\textbf{E}\Big[\Big|\int_0^T\int_0^{X(s-)}f(s)\widetilde{N}_\eta(ds,dz)\Big|^{2p}\Big]
 	\leq C\Big(\eta\cdot\int_0^T|f(s)|^2ds\Big)^p
 	+C\cdot \eta\cdot\int_0^T\big|f(s) \big|^{2p}ds.
 \end{equation} 

 \end{proposition}
 
 The above proposition will be repeatedly applied to the stochastic integral in (\ref{eqn.Jn}) to obtain the necessary moment estimates. 
 In our setting the stochastic process $\omega $ in \eqref{2pine} is hence given by the rescaled volatility process(es) $V^{(n)}$. 
 To apply this proposition in the sequel, we thus need to verify that the rescaled volatility processes are $L^p$-bounded. 
 This is achieved by the next lemma.

 \begin{lemma} \label{V2p}
 For any $p\geq 0$ and $T\geq 0$, the stochastic integral term $J^{(n)} $ defined in (\ref{eqn.Jn}) satisfy 
 \beqnn
 \sup_{n\geq 1}\sup_{t\in[0,T]} \mathbf{E}\Big[\big|J^{(n)}(t)\big|^{2p}\Big]<\infty
 \quad \mbox{and}\quad 
  \sup_{n\geq 1}\sup_{t\in[0,T]} \mathbf{E}\Big[\big|V^{(n)}(t)\big|^{2p}\Big]<\infty. 
 \eeqnn
 
 \end{lemma}
 \proof 
 By Jensen's inequality it suffices to consider the cases $2p=2^k$ with  $ k\in \{0, 1, 2, ...\}$. We proceed by induction, starting with the case $k=0$. 
 
 By Corollary~\ref{Cor.ConvergeVI}, the first two functions on the right side of (\ref{decomV}) are uniformly bounded, i.e., 
 \beqlb\label{eqn.Upperbound.01}
 \sup_{n\geq 1}\sup_{t\geq 0} V^{(n)}_0(t)  
 \leq \sup_{n\geq 1}  V^{(n)}_0(0) <\infty
 \quad \mbox{and}\quad 
 \sup_{n\geq 1}\sup_{t\geq 0} I^{(n)}(t)
 \leq \sup_{n\geq 1}  I^{(n)}(\infty)<\infty. 
 \eeqlb
 To establish the uniform upper bound for $\{V^{(n)}\}_{n\geq 1}$,  it hence suffices to consider the stochastic integral processes $\{J^{(n)}\}_{n\geq 1}$. We note that these are not (local) martingales because of the time-dependence of the integrands. 
 
 Nonetheless, we can still compute their expectations by introducing an auxiliary stochastic process. Specifically, for any $t\in[0,T]$ and $n\geq 1$, we define the process
 \beqnn
 J^{(n)}_{t}(r) := \int_0^r \int_0^{V^{(n)}(s-)}   \frac{\zeta_n}{n^{2\alpha-1}}\cdot R_n\big(n(t-s)\big)  \widetilde{N}^{(n)}_{1}(ds,dz),
 \quad r\in [0,t].
 \eeqnn
 The process $ J^{(n)}_{t} $ is a martingale on $[0,t]$; by construction $J^{(n)}_{t}(0)=0$ and $J^{(n)}_{t}(t)=J^{(n)} (t)$. 
 As a result,  
 \beqlb  \label{expectation}
 \mathbf{E}\Big[ \big|J^{(n)} (t) \big|^{2^k}\Big]= \mathbf{E}\Big[\big|J^{(n)}_{t}(t)\big|^{2^k}\Big] . 
 \eeqlb
 
 When $k=0$, we take expectations on both sides of (\ref{rescaledV}) and then use (\ref{eqn.Upperbound.01}) to get that 
 \beqlb \label{eqn.Upperbound.02}
 \mathbf{E}\Big[\big|V^{(n)}(t)\big|\Big]= \mathbf{E}\Big[ V^{(n)}(t) \Big] =   V^{(n)}_0(t)  + I^{(n)}(t)  
 \quad \mbox{and hence}\quad 
 \sup_{n\geq 1}\sup_{t\geq 0} \mathbf{E}\Big[\big|V^{(n)}(t)\big| \Big]\leq C,
 \eeqlb
 for some constant $C>0$ independent of $n$ and $t$. 
 By using the Burkholder-Davis-Gundy inequality and then Jensen's inequality to \eqref{expectation}, there exists a constant $C>0$ that is independent of $n$ and $t$ such that 
 \beqnn
 \mathbf{E}\Big[ \big|J^{(n)} (t) \big| \Big]= \mathbf{E}\Big[\big|J^{(n)}_{t}(t)\big| \Big]
 \ar\leq\ar C \mathbf{E}\Big[ \Big|\int_0^t \int_0^{V^{(n)}(s-)}  \frac{\big|R_n\big(n(t-s)\big)\big|^2}{n^{4\alpha-2}}   N^{(n)}_{1}(ds,dz)\Big|^{1/2}\Big]\cr
 \ar\leq\ar C  \Big(\mathbf{E}\Big[\int_0^t \int_0^{V^{(n)}(s-)}  \frac{\big|R_n\big(n(t-s)\big)\big|^2}{n^{4\alpha-2}}  N^{(n)}_{1}(ds,dz)\Big]\Big)^{1/2} \cr
 \ar=\ar C  \Big(\int_0^t \mathbf{E}\big[V^{(n)}(s)\big] \frac{\big|R_n\big(n(t-s)\big)\big|^2}{n^{2\alpha-2}}   ds\Big)^{1/2}. 
 \eeqnn
 Applying (\ref{eqn.Upperbound.02}) and Proposition~\ref{boundR} to the last term, we have 
 \beqnn
 \mathbf{E}\Big[ \big|J^{(n)} (t) \big| \Big] \leq C  \Big(\int_0^t  \frac{(1+ns)^{2\alpha-2}}{n^{2\alpha-2}}   ds\Big)^{1/2} \leq C  \Big(\int_0^t   s^{2\alpha-2}    ds\Big)^{1/2} \leq C,
 \eeqnn
 uniformly in $n\geq 1$ and $t\in[0,T]$.  This proves the desired result for $k=0$. 
 To proceed we assume that the desired inequality holds for $2p=2^k$ for some $k \geq 0$ and prove that it holds for $2p=2^{k+1}$.  
 
 First, from the power mean inequality and (\ref{eqn.Upperbound.01}), we see that there exists a constant $C>0$ such that for all $t\geq 0$ and $ n\geq 1$, 
 \beqnn
 \mathbf{E}\Big[\big|V^{(n)}(t)\big|^{2p}\Big]\leq C\cdot \Big(1
 +\mathbf{E}\Big[\big|J^{(n)}(t)\big|^{2p}\Big]\Big).
 \eeqnn
 Furthermore, the induction hypotheses allows us to apply  Proposition~\ref{2pine}  to $\mathbf{E}\big[|J^{(n)} (t)|^{2p}\big] $ to obtain that uniformly in $n\geq 1$ and $t\in[0,T]$,  
 \beqnn
 \mathbf{E}\Big[\big|J^{(n)}(t)\big|^{2p}\Big]
 \ar\leq\ar 
 C\cdot  
 \Big( n^{2\alpha}\int_0^t \Big(\frac{R_n(ns)}{n^{2\alpha-1}}\Big)^2ds\Big)^p+
  C\cdot n^{2\alpha}\int_0^t\Big(\frac{R_n(ns)}{n^{2\alpha-1}}\Big)^{2p}ds.  
 \eeqnn
 
 \begin{itemize}
 \item Using that $\alpha > 1/2$ and the inequality \eqref{ubRH} the first term on the right side of the above inequality can be estimated as follows: 
 \beqnn
  \Big( n^{2\alpha}\int_0^t \Big(\frac{R_n(ns)}{n^{2\alpha-1}}\Big)^2ds\Big)^p
  \leq  C \Big(\int_0^t\frac{(1+ns)^{2\alpha-2}}{n^{2\alpha-2}}ds \Big)^p \leq  C \cdot (1+t)^{2\alpha p}. 
 \eeqnn 
 
 \item Using \eqref{ubRH} again the second term also can be estimated as follows:
 \beqnn
 n^{2\alpha}\int_0^t\Big(\frac{R_n(ns)}{n^{2\alpha-1}}\Big)^{2p}ds \ar\leq\ar C \frac{n^{2\alpha+2p(\alpha-1)} }{n^{2p(2\alpha-1)}} \int_0^t
 \big(1/n+s\big)^{2p(\alpha-1)}ds \cr
 \ar=\ar \frac{ C\cdot n^{2\alpha(1-p)}}{2p(\alpha-1)+1}\Big(
 \big(1/n +t \big)^{2p(\alpha-1)+1}- (1/n)^{2p(\alpha-1)+1}\Big). 
 \eeqnn
 We now distinguish two cases:
 \begin{itemize}
 	\item 
 	If $2p(\alpha-1)+1<0$, then using that $\alpha>1/2$ and $p\geq 1$ (induction hypothesis) we see that, 
 	\beqnn
 	n^{2\alpha}\int_0^t\Big(\frac{R_n(ns)}{n^{2\alpha-1}}\Big)^{2p}ds  \leq 
 	C\cdot n^{2\alpha(1-p)} \cdot \big(1/n\big)^{2p(\alpha-1)+1}=C\cdot n^{-(2p-1)(2\alpha-1)}\leq C. 
 	\eeqnn
 	\item If $2p(\alpha-1)+1>0$, then again using that $p=2^k\geq 1$,  
 	\beqnn
 	n^{2\alpha}\int_0^t\Big(\frac{R_n(ns)}{n^{2\alpha-1}}\Big)^{2p}ds  \leq
 	Cn^{2\alpha(1-p)} \cdot \big(1/n+t\big)^{2p(\alpha-1)+1}
 	\leq C(1+t)^{2\alpha p}. 
 	\eeqnn
 \end{itemize}
 \end{itemize}  
 Combing the above estimates shows that 
 \beqnn
 \sup_{n\in\mathbb N}\sup_{t\in[0,T]}\mathbf{E}\Big[\big|V^{(n)}(t)\big|^{2p}\Big] \leq \sup_{t\in[0,T]}C(1+t)^{2\alpha p} \leq C(1+T)^{2\alpha p}<\infty.
 \eeqnn
 \qed
 
 \subsection{$C$-tightness of $\{J^{(n)} \}_{n\geq 1}$}\label{Sec.tight}
 
 In this section we establish the $C$-tightness of  the c\`adl\`ag process sequence $\{J^{(n)} \}_{n\geq 1}$. 
 Because of the existence of jumps, the moments of the increments $\Delta_h J^{(n)}$ cannot be uniformly bounded by a term of the form $C\cdot h^{1+\kappa}$, which prevents us from applying the standard Kolmogorov-Chentsov criterion. 
 Instead, we prove that the sequence meets the conditions presented in Lemma \ref{tightness condition}. 
 
 Notice that $J^{(n)}(0) =0$ a.s., it suffices to identify that the two conditions in Lemma \ref{tightness condition} are satisfied. 
 To prove that the first condition holds it will be convenient to represent our integral processes in terms of the derivative of the resolvent function. The resolvents and their first derivatives are smooth and bounded functions, and satisfy
 \beqnn
 R_n \big(n(t-s)\big)= R_n(0) + \int_s^t n\cdot R'_n \big(n(r-s)\big)\, dr,\quad n\geq 1,\ t\geq s\geq 0. 
 \eeqnn 
 As a  result of (\ref{eqn.Rn}) and so
 $R_n(0)=\alpha \sigma  \zeta_n$, we can write $ J^{(n)}  $ as 
 \beqlb\label{eqn.SplitJn}
 J^{(n)}(t)=J^{(n)}_1 (t) +  J^{(n)}_2 (t),\quad t\geq 0,
 \eeqlb
 with
 \beqlb
  J^{(n)}_1 (t) \ar:=\ar \int_0^t \int_0^{V^{(n)}(s-)}\frac{\alpha \sigma  \zeta_n}{n^{2\alpha-1}} \widetilde{N}^{(n)}_{1}(ds,dz), \label{eqn.Jn1}\\
 J^{(n)}_2 (t) \ar:=\ar \int_0^t \int_0^s \int_0^{V^{(n)}(r-)}\frac{R'_n\big(n(s-r)\big)}{  n^{2\alpha-2} }  \widetilde{N}^{(n)}_{1}(dr,dz) ds . \label{eqn.Jn2}
 \eeqlb 
 
 {The following lemma proves that condition (1) in Lemma \ref{tightness condition} is satisfied and hence that the sequence of linear interpolations $\{J^{(n)}_\theta \}_{n\geq 1}$ defined as in (\ref{interpolation}) is asymptotically indistinguishable from $\{J^{(n)}  \}_{n\geq 1}$.  }
 
 \begin{lemma} \label{Imtight}
 For any $\theta > 2$, the process sequence  $\{J^{(n)}\}_{n\geq 1}$ satisfies as $n\to\infty$,
 \beqnn
 	\sup_{k=0,1,\cdots,[Tn^\theta]} \sup_{h\in[0,1/n^{\theta}]} \big|\Delta_h J^{(n)}  (k/n^\theta)\big|  \overset{\rm p}\to 0 . 
 \eeqnn

 \end{lemma}
 \proof
 In terms of the processes $J^{(n)}_{i}$, $i=1,2$, introduced in (\ref{eqn.Jn1})-(\ref{eqn.Jn2}) it thus suffices to prove that
 \beqnn
 \sup_{k=0,1,\cdots,[Tn^\theta]} \sup_{h\in[0,1/n^{\theta}]} \big|\Delta_h J^{(n)}_{i} (k/n^\theta) \big|  \overset{\rm p}\to 0  .
 \eeqnn
 
 \medskip
 
 {\bf Case $i=1$.} We start with the first process that does not involve the derivative of the resolvent. For any $\eta>0$, by Chebyshev's inequality with 
 \beqlb\label{eqn.002}
 p>\frac{2\alpha+\theta}{2\alpha-1} \geq \theta+2, 
 \eeqlb
 we have for any $n\geq 1$,
 \beqlb\label{eqn.001}
 \mathbf{P}\Big( \sup_{k=0,1,\cdots,[Tn^\theta]} \sup_{h\in[0,1/n^{\theta}]} \big|\Delta_h J^{(n)}_1 (k/n^\theta) \big| \geq \eta  \Big)
 \ar\leq\ar  \sum_{k=0}^{[Tn^\theta]} \frac{1}{\eta^{4p}} \mathbf{E}\Big[\sup_{h\in[0,1/n^{\theta}]}  \big|\Delta_h J^{(n)}_1 (k/n^\theta) \big|^{4p}\Big]
 \eeqlb
 Applying the Burkholder-Davis-Gundy inequality, together with the fact that $ \zeta_n \sim 1$, to the last expectation gives that for some constant $C>0$ independent of $n$ and $k$,  
 \beqnn
 \mathbf{E}\Big[\sup_{h\in[0,1/n^{\theta}]}  \big|\Delta_h J^{(n)}_1 (k/n^\theta) \big|^{4p}\Big]
 \leq C \cdot \mathbf{E}\Big[ \Big|\int_{k/n^\theta} ^{(k+1)/n^\theta} \int_0^{V^{(n)}(s-)}\frac{N^{(n)}_{1}(ds,dz) }{ n^{4\alpha-2}}\Big|^{2p}\Big]. 
 \eeqnn
 Expressing the Poisson random measure $N^{(n)}_{1}(ds,dz)$ as the sum of the compensated measure $\widetilde{N}^{(n)}_{1}(ds,dz)$ and the compensator $n^{2\alpha}\cdot dsdz$, and then using the power mean inequality,  the expectation on the right-hand side of the above inequality can be bounded by 
 \beqnn
 2^{2p}\cdot\mathbf{E}\Big[ \Big|\int_{k/n^\theta} ^{(k+1)/n^\theta} \int_0^{V^{(n)}(s-)}\frac{\widetilde{N}^{(n)}_{1}(ds,dz)}{ n^{4\alpha-2}} \Big|^{2p}\Big] 
 +   2^{2p}\cdot \mathbf{E}\Big [\Big|\int_{k/n^\theta}^{(k+1)/n^\theta}  \frac{V^{(n)}(s)}{n^{2\alpha-2}} ds \Big|^{2p} \Big] .
 \eeqnn
 Using Proposition~\ref{2pine} and Lemma~\ref{V2p} to the first expectation, it can be bounded by
 \beqnn
 C\cdot \int_{k/n^\theta}^{(k+1)/n^\theta}\frac{n^{2\alpha}ds}{n^{4p(2\alpha-1)}} 
 + C \cdot \Big(\int_{k/n^\theta}^{(k+1)/n^\theta}\frac{n^{2\alpha} ds}{n^{8\alpha-4}} \Big)^p 
 \leq C\cdot \Big( n^{2\alpha-\theta-4p(2\alpha-1) }  +   n^{p(4-6\alpha-\theta)} \Big),
 \eeqnn
 for some constant $C>0$ independent of $n$ and $k$. 
 Applying H\"older's inequality to the second expectation and then using Lemma~\ref{V2p}, it can be uniformly bounded by 
 \beqnn
 n^{-2p(2\alpha-2)-\theta (2p-1)}\int_{k/n^\theta}^{(k+1)/n^\theta}   \mathbf{E}\Big [\big|V^{(n)}(s)\big|^{2p} \Big]  ds \leq C\cdot  n^{-2p(\theta+2\alpha-2) }.
 \eeqnn
 Combining all preceding estimates together and then taking them back into (\ref{eqn.001}), 
 \beqnn
  \mathbf{P}\Big( \sup_{k=0,1,\cdots,[Tn^\theta]} \sup_{h\in[0,1/n^{\theta}]} \big|\Delta_h J^{(n)}_1 (k/n^\theta) \big| \geq \eta  \Big)
  \ar\leq\ar \frac{C}{\eta^{4p}} \Big(  n^{2\alpha -4p(2\alpha-1) }  +   n^{p(4-6\alpha-\theta)+\theta } + n^{\theta- 2p(\theta+2\alpha-2)} \Big) . 
 \eeqnn 
 By using the inequality (\ref{eqn.002}) as well as the facts that $\alpha\in (1/2,1]$ and $\theta > 2$, we can show that all powers in the last term are negative, thereby showing that as $ n\to\infty$,
 \beqnn
  \mathbf{P}\Big( \sup_{k=0,1,\cdots,[Tn^\theta]} \sup_{h\in[0,1/n^{\theta}]} \big|\Delta_h J^{(n)}_1 (k/n^\theta) \big| \geq \eta  \Big) \to 0.
 \eeqnn
 
 \medskip
 {\bf Case $i=2$.} Let us now focus on the processes $\{J^{(n)}_2\}_{n\geq 1}$. 
 For each $\eta>0$, we use Chebyshev's inequality again to get 
 \beqlb\label{eqn.003}
 \mathbf{P}\Big( \sup_{k=0,1,\cdots,[Tn^\theta]} \sup_{h\in[0,1/n^{\theta}]}\big|\Delta_h J^{(n)}_2(k/n^\theta)\big| \geq \eta \Big) \leq \frac{1}{\eta}\cdot \mathbf{E}\Big[ \sup_{k=0,1,\cdots,[Tn^\theta]} \sup_{h\in[0,1/n^{\theta}]}\big|\Delta_h J^{(n)}_2(k/n^\theta)\big|  \Big].
 \eeqlb
 Since $\zeta_n \sim 1$,  by (\ref{eqn.Jn2}) we have that 
 \beqnn
 \big|\Delta_h J^{(n)}_2(t)\big|
  \leq h\cdot  \sup_{r\in[0,T]} \Big| \int_0^r \int_0^{V^{(n)}(s-)}\frac{R'_n(n(r-s)) }{  n^{2\alpha-2} }  \widetilde{N}^{(n)}_{1}(ds,dz) \Big|,
 \eeqnn
 uniformly in $t\in[0,T]$ and $n\geq 1$, which yields that
 \beqnn
 \sup_{k=0,1,\cdots,[Tn^\theta]} \sup_{h\in[0,1/n^{\theta}]}\big|\Delta_h J^{(n)}_2(k/n^\theta)\big| 
 \ar\leq\ar  \frac{C}{n^\theta}\cdot \sup_{t\in[0,T]} \Big| \int_0^t \int_0^{V^{(n)}(s-)}\frac{R'_n(n(t-s)) }{  n^{2\alpha-2} }  \widetilde{N}^{(n)}_{1}(ds,dz) \Big| \cr
 \ar\leq\ar  \frac{C}{n^\theta}\cdot \sup_{t\in[0,T]}  \int_0^t \int_0^{V^{(n)}(s-)}\frac{|R'_n(n(t-s))| }{  n^{2\alpha-2} }  N^{(n)}_{1}(ds,dz)  \cr
 \ar\ar + \frac{C}{n^{\theta-2}}\cdot \sup_{t\in[0,T]}   \int_0^t  V^{(n)}(s-) \big|R'_n(n(t-s))\big|  ds  .
 \eeqnn 
 Using Proposition \ref{propG} and the fact that $\alpha < 1$ shows that $ \sup_{n\geq 1}\sup_{t\geq 0}|R'_n(nt)|<\infty$  and hence that 
 \beqnn
  \sup_{k=0,1,\cdots,[Tn^\theta]} \sup_{h\in[0,1/n^{\theta}]}\big|\Delta_h J^{(n)}_2(k/n^\theta)\big| 
  \ar\leq\ar \frac{C}{n^\theta}  \int_0^T \int_0^{V^{(n)}(s-)}\frac{N^{(n)}_{1}(ds,dz) }{  n^{2\alpha-2} }  + \frac{C}{n^{\theta-2}}  \int_0^T  V^{(n)}(s)    ds  .
 \eeqnn
 Taking expectations on both sides of this inequality and then using Lemma~\ref{V2p}, 
 \beqnn
 \mathbf{E}\Big[ \sup_{k=0,1,\cdots,[Tn^\theta]} \sup_{h\in[0,1/n^{\theta}]}\big|\Delta_h J^{(n)}_2(k/n^\theta)\big| \Big] 
 \leq C\cdot n^{2-\theta}.
 \eeqnn
 Taking this back into (\ref{eqn.003}) and using the fact that $\theta>2$, we have as $n\to\infty$,
 \beqnn
 \mathbf{P}\Big( \sup_{k=0,1,\cdots,[Tn^\theta]} \sup_{h\in[0,1/n^{\theta}]}\big|\Delta_h J^{(n)}_2(k/n^\theta)\big| \geq \eta \Big) \to 0. 
 \eeqnn 
 \qed

 The next lemma proves that the sequence $\{J^{(n)}\}_{n\geq 1}$ satisfies condition (2) in Lemma \ref{tightness condition}. 
 For the lemma to hold, it is important to have the  locally uniform moment estimates  of all orders established in Lemma \ref{V2p}.

 \begin{lemma}
 For each $T\geq 0$,  $p>\frac{1}{2}$ and $\delta\in(0,1)$ there exists a constant $C>0$ such that for any $h\in(0,1)$ and $n\geq 1$ the following moment estimate holds:
 \beqlb\label{eqn.005}
 \sup_{t\in[0,T]} \mathbf{E}\Big[\big|\Delta_h J^{(n)}(t)\big|^{2p}\Big] \leq C\Big( h^{p(2\alpha-1)}+ \frac{h^{\frac{\delta}{2-\alpha}}}{n^{2p(2\alpha-1)+\frac{1-\alpha}{2-\alpha}-2\alpha}}\Big).
 \eeqlb
 Moreover, we can choose the two constants $p$ and $\delta$ such that condition (2) in Lemma \ref{tightness condition} is satisfied.
 \end{lemma}
 \proof 
 For any $n\geq 1 $ and $t\in[0,T]$, we can express the increment $\Delta_h J^{(n)}(t)$ as the summation of the following two terms
 \beqnn
 \epsilon_1^{(n)}(t,h)\ar:=\ar \int_t^{t+h}  \int_0^{V^{(n)}(s-)} \frac{1}{ n^{2\alpha-1}} R_n(n(t+h-s)) \widetilde{N}^{(n)}_1(ds,dz), \cr
 \epsilon_2^{(n)}(t,h)\ar:=\ar \int_0^t  \int_0^{V^{(n)}(s-)} \frac{1}{n^{2\alpha-1}} \big(R_n(n(t+h-s))- R_n(n(t-s))\big)  \widetilde{N}^{(n)}_1(ds,dz).
 \eeqnn                                    
 By the power mean inequality, we have 
 \beqnn
  \sup_{t\in[0,T]} \mathbf{E}\Big[\big|\Delta_h J^{(n)}(t)\big|^{2p}\Big] 
  \leq 2^{2p} \cdot  \sup_{t\in[0,T]} \mathbf{E}\Big[\big|\epsilon_1^{(n)}(t,h)\big|^{2p}\Big]   + 2^{2p} \cdot  \sup_{t\in[0,T]} \mathbf{E}\Big[\big|\epsilon_2^{(n)}(t,h)\big|^{2p}\Big] .
 \eeqnn
 In the next two steps, we  establish the desired uniform upper bound for the two supremums on the right of this inequality, i.e.,
 \beqlb\label{eqn.006}
 \sup_{t\in[0,T]} \mathbf{E}\Big[\big|\epsilon_i^{(n)}(t,h)\big|^{2p}\Big] \leq  C\Big( h^{p(2\alpha-1)}+ \frac{h^{\delta/(2-\alpha)}}{n^{2p(2\alpha-1)+\frac{1-\alpha}{2-\alpha}-2\alpha}}\Big), 
 \quad i=1,2.
 \eeqlb
 
 \medskip
 
 {\bf Case $\epsilon_1^{(n)}$.}   
 The moment estimate of $\epsilon_1^{(n)}(t,h)$ can be obtained as follows. 
  Applications of Proposition~\ref{2pine} together with Lemma~\ref{V2p} and the fact that $\zeta_n\sim 1$ yield that for some constant $C>0$ that is independent of $n$, $t$ and $h$ such that 
 \beqnn
  \mathbf{E}\Big[\big|\epsilon_1^{(n)}(t,h)\big|^{2p}\Big] 
  \ar\leq\ar C\Big|\int_0^h n^{2\alpha}\cdot \Big(\frac{R_n(ns)}{n^{2\alpha-1}}\Big)^2ds\Big|^p+ C\Big|\int_0^h n^{2\alpha}\cdot \Big(\frac{R_n(ns)}{n^{2\alpha-1}}\Big)^{2p}ds\Big|\cr
  \ar\leq\ar C\Big|\int_0^h  \big|n^{1-\alpha}R_n(ns)  \big|^2 ds  \Big|^p
  + C\cdot n^{2\alpha-2p(2\alpha-1)}\int_0^h  \big(R_n(ns)\big)^{2p} ds.
 \eeqnn
 The upper bound on the resolvent $R_n$ established in \eqref{ubRH} and the fact that $\alpha\in(1/2,1)$ yields that uniformly in $n\geq 1$ and $h\in(0,1)$, 
 \beqnn
 \Big|\int_0^h  \big|n^{1-\alpha}R_n(ns)  \big|^2 ds  \Big|^p\leq C \cdot \Big|\int_0^h s^{2\alpha-2}ds\Big|^p \leq C h^{p(2\alpha-1)}.    
 \eeqnn 
 To bound the second term we use the facts that $\frac{1}{2-\alpha}<2p$ and $\alpha-1<0$ from which we obtain that $ \big|R_n(ns)\big|^{2p} \leq C\cdot (1+ns)^{\frac{\alpha-1}{2-\alpha}} \leq C\cdot ( ns)^{\frac{\alpha-1}{2-\alpha}}$ uniformly in $s\geq 0$ and $n\geq 1$, which yields that 
 \beqnn
 \frac{1}{n^{2p(2\alpha-1)-2\alpha}}	\int_0^h  \big|R_n(ns)\big|^{2p} ds \leq C\cdot \frac{h^{\frac{1}{2-\alpha}} }{n^{2p(2\alpha-1)-2\alpha +\frac{1-\alpha}{2-\alpha}}}  . 
 \eeqnn 
 Therefore, the inequality (\ref{eqn.006}) holds  uniformly in  $ n\geq 1$ and $h\in (0,1)$.  
 
 \medskip
 
 {\bf Case $\epsilon_1^{(n)}$.} We now establish the uniform upper-bound (\ref{eqn.006}) for $i=2$. 
 The equality  
 \beqnn
 R_n\big(n(t+h-s)\big)- R_n\big(n(t-s)\big)=\int_{t-s}^{t+h-s} n R'_n(nr) \, dr
 \eeqnn
 allows us to write $\epsilon_2^{(n)}(t,h)$ into 
 \beqnn
 \epsilon_2^{(n)}(t,h) = \int_0^t  \int_0^{V^{(n)}(s-)} 
 \Big(\int_{t-s}^{t+h-s} 
 \frac{ R'_n(nr)}{ n^{2\alpha-2}} 
  \, dr\Big)
 \widetilde{N}^{(n)}_1(ds,dz).
 \eeqnn 
 Using Proposition~\ref{2pine} and the fact that $\zeta_n\sim 1$ again to $\mathbf{E}\big[ |\epsilon_2^{(n)}(t,h) |^{2p}\big]$, it can be bounded uniformly in $n\geq 1$, $t\in[0,T]$ and $h\in (0,1)$ by 
 \beqlb\label{eqn.007}
 C \Big| \int_0^t  \Big(\int_{s}^{s+h} 
 n^{2-\alpha} R'_n(nr) \,
  dr\Big)^2 ds  \Big|^p
 +\frac{C}{n^{2p(2\alpha-1)-2\alpha}}  \int_0^t \Big|\int_{s}^{s+h}n\cdot R'_n(nr)\, dr\Big|^{2p} ds.
 \eeqlb
 It follows from \eqref{ubGH} that $ n^{2-\alpha} R'_n(nr) \leq C\cdot r^{\alpha-2}$ uniformly in $r>0$. 
 Hence, by the power mean inequality the first term can be bounded by
 \beqnn
  C \Big| \int_0^t  \Big(\int_{s}^{s+h} r^{\alpha-2}dr\Big)^2 ds  \Big|^p
  \ar\leq\ar C \Big| \int_0^h  \Big(\int_{s}^{s+h} r^{\alpha-2}dr\Big)^2 ds  \Big|^p+ C\Big| \int_h^t  \Big(\int_{s}^{s+h} r^{\alpha-2}dr\Big)^2 ds  \Big|^p \cr
  \ar\leq\ar C \Big| \int_0^h s^{2\alpha-2}  ds  \Big|^p+ C\Big| \int_h^t  \big|s^{\alpha-2}\cdot h\big|^2 ds  \Big|^p
  \leq  C \cdot h^{p(2\alpha-1)}. 
 \eeqnn
 Here the constant $C>0$ is independent of $t$ and $h$. 
 We turn to consider the second term in (\ref{eqn.007}). 
 By Proposition~\ref{boundR}, 
 \beqnn
 \sup_{n\geq 1}\sup_{s\geq 0} \Big|\int_{s}^{s+h}n\cdot R'_n(nr)dr \Big|\leq \sup_{n\geq 1} R_n(0) <\infty,
 \eeqnn 
 which, together with the facts that $\frac{1}{2-\alpha}<2p$ and $\alpha-1<0$, yields that uniformly in $n\geq 1$, $t\in[0,T]$ and $h\in(0,1)$, 
 \beqnn
  \int_0^t \Big|\int_{s}^{s+h}n\cdot R'_n(nr)dr\Big|^{2p} ds 
  \ar\leq\ar C\cdot \int_0^t \Big|\int_{s}^{s+h}n\cdot R'_n(nr)dr\Big|^{\frac{1}{2-\alpha}} ds \cr
  \ar=\ar C\cdot n^{-\frac{1-\alpha}{2-\alpha}} \int_0^t \Big|\int_{s}^{s+h}n^{2-\alpha}\cdot R'_n(nr)dr\Big|^{\frac{1}{2-\alpha}} ds. 
 \eeqnn
 By the facts that $\alpha-1<0$ and  $ n^{2-\alpha} R'_n(nr) \leq C\cdot r^{\alpha-2}$ uniformly in $r>0$, the last integral can be bounded uniformly in $n\geq 1$, $t\in[0,T]$ and $h\in(0,1)$ by
 \beqnn
  \int_0^t \Big|\int_{s}^{s+h}r^{\alpha-2} dr\Big|^{\frac{1}{2-\alpha}} ds
  \ar=\ar \int_0^h \Big|\int_{s}^{s+h}r^{\alpha-2}dr\Big|^{\frac{1}{2-\alpha}} ds 
  + \int_h^t \Big|\int_{s}^{s+h}r^{\alpha-2}dr\Big|^{\frac{1}{2-\alpha}} ds \cr
  \ar\leq\ar C\int_0^h s^{\frac{\alpha-1}{2-\alpha}} ds + C\cdot h^{\frac{1}{2-\alpha}} \int_h^t |s^{\alpha-2}|^{\frac{1}{2-\alpha}}ds \\
  \ar\leq\ar C\cdot h^{\frac{1}{2-\alpha}}(1+\log h) 
  \leq  C\cdot h^{\frac{\delta}{2-\alpha}}, 
 \eeqnn
 for any $\delta\in(0,1)$. 
 Taking all preceding estimates back into (\ref{eqn.007}) induces that  the inequality (\ref{eqn.006}) holds  uniformly in  $ n\geq 1$ and $h\in (0,1)$. 
 
 \medskip
 
 Finally, we show that condition (2) in Lemma \ref{tightness condition} can be satisfied by choosing 
 \beqnn
 p>\frac{1+\theta}{2\alpha-1}
 \geq \frac{1-\theta(\alpha+\delta-2)}{2(2-\alpha)(2\alpha-1)} +\frac{1}{2}.
 \eeqnn
 Indeed, the moment estimate (\ref{eqn.005}) is of the form \eqref{condition1} with
 \beqnn
 m=2,\quad  a_1 = 0, \quad b_1=p(2\alpha-1), \quad a_2=2p(2\alpha-1)+\frac{1-\alpha}{2-\alpha} -2\alpha, \quad b_2=\frac{\delta}{2-\alpha}.
 \eeqnn
 It is obvious that  $b_1+a_1/\theta=b_1 > 1$.
 Moreover,
 \beqnn
 \theta b_2+a_2-\theta 
 \ar=\ar\frac{\theta\cdot \delta}{2-\alpha}+2p(2\alpha-1)+\frac{1-\alpha}{2-\alpha} -2\alpha-\theta 
 = \frac{\theta(\alpha+\delta-2)-1}{2-\alpha}   +(2p-1)(2\alpha-1) >0 , 
 \eeqnn
 which yields that $b_2+a_2/\theta>1$.  
 \qed
 
 \begin{corollary}\label{Corollary.JTight}
 The sequence $\{J^{(n)}\}_{n\geq 1} $ is $C$-tight. 
 \end{corollary}
 \proof 
 The preceding two lemmas show that the sequence $\{J^{(n)}\}_{n\geq 1} $ in $\mathbf{D}(\mathbb{R}_+;\mathbb{R})$ satisfies  all conditions in Lemma \ref{tightness condition} and hence is $C$-tight.
 \qed

 \subsection{Convergence of $\{ J^{(n)} \}_{n\geq 1}$}\label{Sec.conv}
 


As the integrands in the definition of the processes $\{ J^{(n)} \}_{n\geq 1}$ depend on the time variable, we cannot establish their weak convergence as in \cite{HorstXu2022} by appealing the standard convergence results established in, e.g.~\cite{KurtzProtter1996}.  Instead, we identify the weak limit by identifying the weak limit the sequence of the integrated processes and then differentiate that limit. More precisely, the weak convergence of the sequence $\{J^{(n)} \}_{n\geq 1}$ in $\mathbf{D}(\mathbb{R}_+;\mathbb{R})$ implies the weak convergence in $L^1(\mathbb{R}_+;\mathbb R)$, which in turns implies the weak convergence of the integrated processes 
 \beqnn
 \mathcal{I}_{J^{(n)}}(t):= \int_0^t  J^{(n)}(s)ds ,\quad t\geq 0,\ n\geq 1.
 \eeqnn
 That is, as $n\to\infty$, 
 \beqlb\label{eqn.0019}
 J^{(n)} \overset{\rm d}\to J_*\quad  \mbox{in }\mathbf{D}(\mathbb{R}_+;\mathbb{R})
 \quad \mbox{implies} \quad 
 \mathcal{I}_{J^{(n)}}(t) \overset{\rm d}\to \mathcal{I}_{J_*}(t):= \int_0^t J_*(s)ds\quad  \mbox{ in }\mathbf{D}(\mathbb{R}_+;\mathbb{R}). 
 \eeqlb 
 Since the (unknown) limiting process $J_*$ is continuous because of the previously established $C$-tightness, it can be identified by differentiating the limit of the integrated process. 
The following proposition confirms our intuition and identifies the weak limit of the integral processes.

 \begin{proposition} \label{Prop.3.13}
 Any accumulation point $(V_*,J_*) \in \mathbf{C}(\mathbb{R}_+;\mathbb{R}_+\times \mathbb R)$ of the sequence $\{ (V^{(n)},J^{(n)} )  \}_{n\geq 1}$ is a weak solution to the stochastic equation 
 \beqlb\label{eqn.0016}
 J_*(t)=
 \int_0^t \frac{1}{b} f^{\alpha,\gamma}(t-s) \sqrt{V_*(s)} dB(s),\quad t\geq 0.
 \eeqlb
 \end{proposition}
 \proof The sequences $\{ V^{(n)}_0  \}_{n\geq 1}$, $\{ I^{(n)}\}_{n\geq 1}$ and $\{J^{(n)}\}_{n\geq 1} $ are $C$-tight. 
 By Corollary~3.33(a) in \cite[p.353]{JacodShiryaev2003} and \eqref{decomV}, the sequence $\{V^{(n)}\}_{n\geq 1}$ is $C$-tight and hence  $\{ (V^{(n)}, J^{(n)}) \}_{n\geq 1}$ is also $C$-tight. 
Let $ ( V_*,J_*) \in \mathbf{C}(\mathbb{R}_+;\mathbb{R}_+\times \mathbb{R})$ be an accumulation point. W.l.o.g.~we may assume that as $n\to\infty$,
 \beqlb\label{eqn.009}
 (V^{(n)}, J^{(n)})\overset{\rm d}\to ( V_*,J_*)  \quad \mbox{ in }\mathbf{D}(\mathbb{R}_+;\mathbb{R}_+\times \mathbb{R}).
 \eeqlb 
 
 To prove that the limit processe $(V_*,J_*)$ solves (\ref{eqn.0016}),
 we first identify the weak limit of the integrated processes $\{\mathcal{I}_{J^{(n)}} \}_{n\geq 1}$ and then identify the desired limit of the integrand processes. We proceed in several steps.
 
 \medskip
 
 {\bf Step 1.}  We first bring the integrated process $\mathcal{I}_{J^{(n)}}$ into a more convenient form. To this end, we introduce the martingale
 \beqlb\label{MaringaleMn}
  M^{(n)}(t):=
  \int_0^t\int_0^{V^{(n)}(s-)}n^{-\alpha}\tilde{N}_1^{(n)}(ds,dz),\quad t\geq 0,
 \eeqlb
 in terms of which we have that
 \beqnn
  J^{(n)}(t)
  =  \int_0^t   n^{1-\alpha} R_n\big(n(t-s)\big)dM^{(n)}(s),
  \quad t\geq 0.
 \eeqnn
 Integrating both side over $[0,t]$ and then applying the stochastic Fubini theorem \cite[Theorem~2.6]{Walsh1986}, 
 \beqlb\label{eqn.0010}
 \mathcal{I}_{J^{(n)}}(t)
 =  \int_0^t  \int_0^r   n^{1-\alpha} R_n\big(n(r-s)\big)dM^{(n)}(s)dr 
 =  \int_0^t  n^{1-\alpha} R_n\big(n(t-s)\big) M^{(n)}(s)ds.
 \eeqlb
 
 \medskip
 
 {\bf Step 2.}  Next, we establish the weak convergence of the sequence $\{M^{(n)}  \}_{n\geq 1}$. 
 The martingale $M^{(n)} $ has predictable quadratic variation
 \beqnn
 \langle M^{(n)}  \rangle_t =\int_0^t V^{(n)}(s)ds,
 \quad t\geq 0. 
 \eeqnn 
 By (\ref{eqn.009}) and the continuous mapping theorem, we have as $n\to\infty$
 \beqnn
  \langle M^{(n)}  \rangle_t \overset{\rm d}\to \int_0^t V_*(s)ds\quad \mbox{in }{\mathbf{C}}(\mathbb{R}_+;\mathbb{R}). 
 \eeqnn
 According to Theorem 4.13 in \cite[p.358]{JacodShiryaev2003}, the tightness of the predictable quadratic variation process guarantees the tightness of the sequence $\{M^{(n)}\}_{n\geq 1}$ and hence the weak convergence along a subsequence to a martingale $M_*$ with predictable quadratic variation 
 \beqnn
 \langle M_*\rangle_t = \int_0^t V_*(s)ds, \quad t\geq 0.
 \eeqnn
 A standard argument shows that $M_*$ is a.s.~continuous.\footnote{We give the complete argument in a slightly more general situation in the proof of Theorem \ref{MainThm.03} below.}  By the martingale representation theorem; see Theorem 7.1 in \cite[p.84]{IkedaWatanabe1989}, there exists a Brownian motion $B$ such that the continuous martingale $M_*$ can be represented as  
 \beqlb\label{eqn.0017}
 M_*(t) = \int_0^t \sqrt{V_*(s)}dB(s),\quad t\geq 0.
 \eeqlb

 To obtain a moment bound of the running supremum of $M_*$ on the time interval we first use the monotone convergence theorem, then the continuous mapping theorem and finally Lemma~\ref{V2p} to get that 
 	\beqnn \label{eqn:0.10}
 	\mathbf{E}\big[ V_*(t) \big] \ar=\ar \lim_{K\to \infty}
 	\mathbf{E}\big[ V_*(t) \wedge K \big] 
 	= \lim_{K\to \infty}\lim_{n\to \infty}
 	\mathbf{E}\big[ V^{(n)}(t) \wedge K \big] 
 	\leq \lim_{n\to \infty}
 	\mathbf{E}\big[ V^{(n)}(t) \big] \leq C ,
 	\eeqnn
 	uniformly in $t\in[0,T]$. This along with the Burkholder-Davis-Gundy inequality  yields that 
 	\beqlb\label{eqn.01}
 	\mathbf{E} \Big[\sup_{t\in[0,T]}\big|M_*(t)\big|^2\Big]
 	\ar\leq\ar C \int_0^{T} \mathbf{E}\big[ V_*(t) \big] ds <\infty.
 	\eeqlb  
 \medskip
 
 {\bf Step 3.}  We now prove the weak convergence of the sequence of integrated processes $\{\mathcal{I}_{J^{(n)}}\}_{n\geq 1}$ in $\mathbf{C}(\mathbb{R}_+;\mathbb{R})$ to 
 	\beqlb\label{eqn.0018}
 	\xi(t):= \int_0^t \frac{1}{b}f^{\alpha,\gamma}(t-s)  M_*(s)ds,\quad t\geq 0.
 	\eeqlb 
In view of the preceding results, we have that
 \beqlb\label{eqn.02}
 (V^{(n)}, J^{(n)}, M^{(n)} ) \overset{\rm d}\to (V_*,J_*,M_*),\quad \mbox{in }\mathbf{D}(\mathbb{R}_+;\mathbb{R}_+\times \mathbb{R}^2). 
 \eeqlb
 By the Skorokhod representation theorem \cite[Theorem I.2.7]{IkedaWatanabe1989}, there exist random variables 
\[
	(\bar V^{(n)}, \bar J^{(n)}, \bar M^{(n)} ) \quad \mbox{ and } \quad (\bar V_*, \bar J_*,\bar M_*) 
\]	
on a possibly different probability space with the same distribution as  $(V^{(n)}, J^{(n)}, M^{(n)} )$, respectively $(V_*,J_*,M_*)$, such that
 \[
 (\bar V^{(n)}, \bar J^{(n)}, \bar M^{(n)} ) \overset{\rm a.s.}\to (\bar V_*, \bar J_*, \bar M_*),\quad \mbox{in }\mathbf{D}(\mathbb{R}_+;\mathbb{R}_+\times \mathbb{R}^2). 
 \]
Continuity of of the processes of $( V_*,J_*, M_*)$ is preserved. In particular,   
%
for any $T\geq 0$,
 \beqlb\label{eqn.008}
 \lim_{n\to\infty}\sup_{t\in[0,T]} \Big( \big| \bar  V^{(n)}(t)-\bar V_*(t) \big| +  \big| \bar J^{(n)}(t)- \bar J_*(t) \big| +\big| \bar M^{(n)}(t)-\bar M_*(t) \big|\Big) \overset{\rm a.s.}= 0 . 
 \eeqlb
 The inequality  \ref{eqn.01} is also preserved. Hence,
 	\beqlb\label{eqn.0111}
 	\mathbf{E} \Big[\sup_{t\in[0,T]}\big| \bar M_*(t)\big|^2\Big]
 	\ar\leq\ar C \int_0^{T} \mathbf{E}\big[ \bar V_*(t) \big] ds <\infty.
 	\eeqlb  
	
 What is not necessarily  preserved is the representation of the processes $J^{(n)}$ as stochastic integrals w.r.t.~to the processes $M^{(n)}$ and hence the {\sl derivation} of the processes $ \mathcal{I}_{J^{(n)}}$.  However, we can {\sl define} processes $\mathcal{I}_{\bar J^{(n)}}$ analogous to \eqref{eqn.0010}  by 
\[
 \mathcal{I}_{\bar J^{(n)}}(t)
  :=  \int_0^t  n^{1-\alpha} R_n\big(n(t-s)\big) \bar M^{(n)}(s)ds.
\]
  Since $\bar M^{(n)}$ is a c\`adl\`ag process and the function $n^{1-\alpha} R_n$ is continuous the above integral is well defined as a Riemann-integral, and hence
 \[
 	\mathcal{I}_{J^{(n)}} \overset{\rm d}= \mathcal{I}_{\bar J^{(n)}}.
 \]

As a result, it suffices to prove that 
 $
  	\mathcal{I}_{\bar J^{(n)}}\overset{\rm d}\to \xi \quad \mbox{in} \quad \mathbf{C}(\mathbb{R}_+;\mathbb{R}).
$
 Since 
 \[
  \big|\mathcal{I}_{\bar J^{(n)}}(t) - \xi(t) \big| \leq \big|\epsilon_1^{(n)}(t)\big|+\big|\epsilon_2^{(n)}(t)\big| 
 \] 
  with
 \beqnn
 \epsilon_1^{(n)}(t)
 \ar:=\ar \int_0^t  \Big( n^{1-\alpha} R_n\big(n(t-s)\big)-  \frac{1}{b}f^{\alpha,\gamma}(t-s)\Big)  \bar M_*(s)ds ,\cr
 \epsilon_2^{(n)}(t)
 \ar:=\ar \int_0^t  n^{1-\alpha} R_n\big(n(t-s)\big)
 \big |\bar M^{(n)}(s)-\bar M_*(s)\big|ds 
 \eeqnn
it is actually enough to prove that the second process converges a.s.~and the first process converges weakly to zero. 

The first result follows from Proposition~\ref{Prop.ConvergenceRn}, which states that the integral process of the rescaled resolvent is uniformly bounded, along with (\ref{eqn.008}) as 
 \beqnn
 \sup_{t\in[0,T]}\big|\epsilon_2^{(n)}(t)\big| \leq  \int_0^T  n^{1-\alpha} R_n(ns)ds \cdot	\sup_{t\in[0,T]}\big |\bar M^{(n)}(t)- \bar M_*(t)\big|   \overset{\rm a.s.}\to 0. 
 \eeqnn
 
 We proceed with the process  $\{\epsilon_1^{(n)} \}_{n\geq 1}$ and first prove its tightness. By a change of variables, 
 \beqnn
 \epsilon_1^{(n)}(t)=\int_0^t  \Big(  n^{1-\alpha} R_n(ns)-  \frac{1}{b}f^{\alpha,\gamma}(s)\Big)  \bar M_*(t-s)ds. 
 \eeqnn
 For any stopping time $\tau\leq T$ and $h\in(0,1)$, by \eqref{eqn.0111} we have 
 \beqnn
 \mathbf{E}\big[|\epsilon_1^{(n)}(\tau+\delta)- \epsilon_1^{(n)}(\tau)|  \big] 
 \ar\leq\ar \mathbf{E} \Big[ \int_0^{\tau+h}  \Big| n^{1-\alpha} R_n(ns)-  \frac{1}{b}f^{\alpha,\gamma}(s)\Big| \big|\bar M_*(\tau+h-s)- \bar M_*(\tau-s)\big|ds\Big]\cr
 \ar\leq\ar  \int_0^{T+1}  \Big( n^{1-\alpha} R_n(ns)+ \frac{1}{b}f^{\alpha,\gamma}(s)\Big) \mathbf{E} \Big[\big| \bar M_*(\tau+h-s)- \bar M_*(\tau-s)\big|\Big] ds\cr
 \ar\leq\ar  \int_0^\infty  \Big( n^{1-\alpha} R_n(ns)+ \frac{1}{b}f^{\alpha,\gamma}(s)\Big)ds\cdot \mathbf{E} \Big[\sup_{t\in[0,T+1]}\big| \bar M_*(t+h)-\bar M_*(t)\big|\Big]\cr
 \ar\leq\ar C\cdot \mathbf{E} \Big[\sup_{t\in[0,T+1]}\big|\bar M_*(t+h)-\bar M_*(t)\big|\Big]
 \eeqnn
 for some constant $C>0$ that is independent of $n$ and $\tau$. 
 Moreover, the continuity of $\bar M_*$ induces that as $h\to 0+$, 
 \beqnn
 \sup_{t\in[0,T+1]}\big|\bar M_*(t+h)-\bar M_*(t)\big| \overset{\rm a.s.}\to 0 
 \quad \mbox{and hence}\quad 
 \mathbf{E} \Big[\sup_{t\in[0,T+1]}\big|\bar M_*(t+h)-\bar M_*(t)\big|\Big] \to 0. 
 \eeqnn 
 The desired tightness of $\{\epsilon_1^{(n)}\}_{n\geq 1}$ thus follows from Aldous's criterion; see \cite{Aldous1978}. 
 
 To prove that $\epsilon_1^{(n)}(t) \overset{\rm a.s.}\to 0$  as $n\to\infty$ for any $t\geq 0$, and hence $\epsilon_1^{(n)}  \to 0$ weakly in  $\mathbf{C}(\mathbb{R}_+;\mathbb{R})$, we introduce two finite measures on $\mathbb{R}_+$:
 \beqnn
 m^{(n)}(ds):= n^{1-\alpha} R_n(ns)ds 
 \quad \mbox{and}\quad 
 m_*(ds):= \frac{1}{b}f^{\alpha,\gamma}(s)ds.
 \eeqnn
  By Proposition~\ref{Prop.ConvergenceRn}, we have $m^{(n)}(ds)\to m_*(ds)$ weakly and hence by the continuity of $\bar M_*$,
 \beqnn
 \int_0^t n^{1-\alpha} R_n(ns) \bar M_*(t-s)ds\ar=\ar \int_0^t \bar M_*(t-s)m^{(n)}(ds) \cr
 \ar\overset{\rm a.s.}\to\ar \int_0^t \bar M_*(t-s)m_*(ds) 
 = \int_0^t \frac{1}{b}f^{\alpha,\gamma}(s)  \bar M_*(t-s)ds,
 \eeqnn
 from which we conclude that $\epsilon_1^{(n)}(t) \overset{\rm a.s.}\to 0$  as $n\to\infty$. 
  We conclude that $\mathcal{I}_{J^{(n)}}
 \to \xi$ weakly in $\mathbf{C}(\mathbb{R}_+;\mathbb{R})$ and hence deduce from Corollary~3.33(b) in \cite[p.353]{JacodShiryaev2003} and \eqref{eqn.02}, that
  \beqlb\label{eqn.0091}
 (V^{(n)}, J^{(n)}, M^{(n)},\mathcal{I}_{J^{(n)}})\overset{\rm d}\to ( V_*,J_*, M_*, \xi) \quad \mbox{ in } \quad \mathbf{D}(\mathbb{R}_+;\mathbb{R}_+\times \mathbb{R}\times \mathbb{R}\times \mathbb{R}).
 \eeqlb

 
  \medskip
 {\bf Step 4.}  It remains to show that (\ref{eqn.0016}) holds.
 By \eqref{eqn.02} and the continuous mapping theorem, we have 
 \beqnn
 (V^{(n)}, J^{(n)}, M^{(n)},\mathcal{I}_{J^{(n)}})\overset{\rm d}\to ( V_*,J_*, M_*, \mathcal{I}_{J^*})  \quad \mbox{ in } \quad \mathbf{D}(\mathbb{R}_+;\mathbb{R}_+\times \mathbb{R}\times \mathbb{R}\times \mathbb{R}),
 \eeqnn
 which, along with \eqref{eqn.0091}, shows that 
 \[
 	\mathcal{I}_{J^*}\overset{\rm a.s.}=\xi.
 \] 
  Similarly as in \eqref{eqn.0010} with $n^{1-\alpha} R_n $ and $M^{(n)} $ replaced by $\frac{1}{b} f^{\alpha,\gamma}$ and $\int_0^\cdot \sqrt{V_*(s)}dB(s)$, we also have 
 \beqnn   
 \int_0^t J_*(s)ds= \mathcal{I}_{J^*}(t)\overset{\rm a.s.}=\xi(t) 
 \ar=\ar \int_0^t \frac{1}{b} f^{\alpha,\gamma}(t-r)  \int_0^r \sqrt{V_*(s)}dB(s)dr\cr
 \ar=\ar \int_0^t\int_0^s \frac{1}{b}f^{\alpha,\gamma}(s-r)\sqrt{V_*(r)}dB(r)ds.
 \eeqnn 
 Since the integrands are continuous, the equality (\ref{eqn.0016}) follows by differentiating both sides of the preceding equality with respect to $t$. 
 \qed 
 
 \subsection{Proof of the main results}\label{Sec.proofs}
 
We are now ready to prove the convergence results stated in Section 2. The convergence of the volatility process follows from the arguments given above. 
 
 \bigskip

 \noindent \textsc{{Proof of Theorem \ref{thm1}}.} 
 As a conclusion of all preceding results, we have proved the following weak convergence in  $\mathbf{D}(\mathbb{R}_+;\mathbb{R}_+^3\times \mathbb{R})$: 
 \beqlb\label{eqn.0020}
 \big( V^{(n)},V_0^{(n)}, I^{(n)}, J^{(n)} \big)  \overset{\rm d}\to \Big( V_*,V_*(0)\cdot(1-F^{\alpha,\gamma}), \frac{a}{b} \cdot F^{\alpha,\gamma},\int_0^\cdot \frac{1}{b} \cdot     f^{\alpha,\gamma}(\cdot-s) \sqrt{V_*(s)} dB(s) \Big). 
 \eeqlb 
 For any $T>0$, by Proposition 2.4 in \cite[p.339]{JacodShiryaev2003} and the continuous mapping theorem we have 
 \beqnn
 \ar\ar\sup_{t\in[0,T]}\Big| V_*(t)-V_*(0)\cdot\big(1-F^{\alpha,\gamma}(t)\big) -\frac{a}{b} \cdot F^{\alpha,\gamma}(t) -\int_0^t \frac{1}{b} \cdot f^{\alpha,\gamma}(t-s) \sqrt{V_*(s)} dB(s) \Big| \cr
 \ar\ar \overset{\rm d}= \lim_{n\to\infty} \big| V^{(n)}(t)-V_0^{(n)}(t)-I^{(n)}(t)-J^{(n)}(t)   \big| \overset{\rm a.s.}=0,
 \eeqnn 
 which shows that 
 \beqnn
 V_*(t)= V_*(0)\cdot\big(1-F^{\alpha,\gamma}(t)\big) +\frac{a}{b} \cdot F^{\alpha,\gamma}(t) +\int_0^t \frac{1}{b} \cdot f^{\alpha,\gamma}(t-s) \sqrt{V_*(s)} dB(s),\quad t\geq 0.
 \eeqnn
 Thus the accumulation point $V_*$ is a weak solution of (\ref{eq1}). 
 The weak uniqueness follows directly from Theorem 6.1 in \cite{JaberLarssonPulido2019}.
 \qed
 
 


The proof of Theorem \ref{thmfractional} can be carried out as in \cite{ElEuchRosenbaum2019b}.  
 We provide an alternative and simpler proof by solving the corresponding Wiener-Hopf equation. 

\bigskip
 
 \noindent \textsc{{Proof of Theorem \ref{thmfractional}}.} We first recall the power function $R^{\alpha,\gamma}$  defined in (\ref{eqn.RML}), which is the resolvent of the Mittag-leffler probability density function $f^{\alpha,\gamma}$; see (\ref{RML}). 
 The stochastic equation (\ref{eqn.FracVol})  can be written as 
 \beqlb \label{eqn.FracVol01}
 V_*(t) = U(t)-  R^{\alpha,\gamma}*V_*(t) , \quad t \geq 0.
 \eeqlb 
 with 
 \beqnn
 U(t):= V_*(0)+ \int_0^t\frac{a}{b} R^{\alpha,\gamma}(s) ds +\int_0^t R^{\alpha,\gamma}(t-s)\cdot\frac{1}{b} \sqrt{V_*(s)}dB_s.
 \eeqnn
 
 In view of Proposition~\ref{eqn.App}, the equation (\ref{eqn.FracVol01}) is equivalent to 
 \beqlb \label{eqn.FracVol02}
 V_*(t)= U(t) - f^{\alpha,\gamma}*U(t),\quad t\geq 0.
 \eeqlb
 Hence it suffices to rewrite this equation into the form (\ref{eq1}). 
 Indeed, applying Fubini's theorem and the stochastic Fubini theorem to $f^{\alpha,\gamma}*U(t)$ shows that
 \beqnn
 f^{\alpha,\gamma}*U(t) \ar=\ar
 V_*(0) \int_0^t f^{\alpha,\gamma}(s)ds + \int_0^tf^{\alpha,\gamma}(t-r)\int_0^r\frac{a}{b}R^{\alpha,\gamma}(s) dsdr\cr
 \ar\ar +\int_0^tf^{\alpha,\gamma}(t-r)\int_0^r R^{\alpha,\gamma}(t-s)\cdot\frac{1}{b} \sqrt{V_*(s)}dB_sdr \cr
 \ar=\ar  V_*(0) F^{\alpha,\gamma} + \int_0^t\frac{a}{b}  R^{\alpha,\gamma}*f^{\alpha,\gamma}(s) ds +\int_0^t R^{\alpha,\gamma}*f^{\alpha,\gamma}(t-s)\cdot\frac{1}{b} \sqrt{V_*(s)}dB_s. 
 \eeqnn
 Taking this back into (\ref{eqn.FracVol02}) and then using the equality $f^{\alpha,\gamma} = R^{\alpha,\gamma} - f^{\alpha,\gamma}*R^{\alpha,\gamma} $ (see (\ref{RML})) yields that
 \beqnn
 \lefteqn{\int_0^t R^{\alpha,\gamma}(t-s)\cdot\frac{1}{b} \sqrt{V_*(s)}dB_s -\int_0^t R^{\alpha,\gamma}*f^{\alpha,\gamma}(t-s)\cdot\frac{1}{b} \sqrt{V_*(s)}dB_s}\ar\ar\cr
 \ar=\ar \int_0^t \big( R^{\alpha,\gamma}(t-s)-R^{\alpha,\gamma}*f^{\alpha,\gamma}(t-s)\big)\cdot\frac{1}{b} \sqrt{V_*(s)}dB_s \\
\ar  = \ar \int_0^t  f^{\alpha,\gamma}(t-s) \cdot\frac{1}{b} \sqrt{V_*(s)}dB_s,
 \eeqnn 
which allows us to rewrite (\ref{eqn.FracVol02}) as (\ref{eq1}). 
 \qed 
 
 \medskip
 
 Having established the convergence of the volatility processes the convergence of the price processes follows from standard arguments. 
 
 \bigskip 
 
  \noindent \textsc{{Proof of Theorem \ref{MainThm.03}}.} 
 For convenience, we represent the rescaled price process $P^{(n)}$ as follows:
  \beqlb\label{eqn.Price01}
  P^{(n)}(t) = P^{(n)}_0+ I_p^{(n)}(t) + J_p^{(n)}(t),\quad t\geq 0,
  \eeqlb
  where $P^{(n)}_0:=n^{-\alpha}\cdot P_{n,0}$ and 
  \beqnn
  I_p^{(n)}(t) \ar:=\ar n^\alpha \int_\mathbb{R} u\nu_n(du)\cdot \int_0^{t} V^{(n)}(s)ds ,\cr
  J_p^{(n)}(t) \ar:=\ar \int_0^{t} \int_{\mathbb R} \int_0^{V^{(n)}(s-)} \frac{u}{ n^\alpha} \widetilde{N}_{n,0}(n\cdot ds, du, n^{2\alpha-1}\cdot dz).
  \eeqnn
   By Assumption~\ref{main.Condition.02} and (\ref{eqn.0020}), the sequence $\{  I_p^{(n)}\}_{n\geq 1}$ converges weakly in $\mathbf{C}(\mathbb{R}_+;\mathbb{R})$ to the process
  \beqnn
  b_p \int_0^t V_*(s)ds,\quad t\geq 0.
  \eeqnn
 
 On the other hand, the process $J^{(n)}_p$ is a martingale with predictable quadratic variation 
  \beqnn
  \langle J^{(n)}_p\rangle_t=\int_\mathbb{R} |u^2|\nu_n(du)\cdot  \int_0^t V^{(n)}(s) ds,
  \eeqnn
  which converges weakly in $\mathbf{C}(\mathbb{R}_+;\mathbb{R})$ to 
  \beqnn
  \sigma_p^2 \int_0^t V_*(s)ds,\quad t\geq 0.
  \eeqnn 
  By \cite[Theorem 4.13, p.358]{JacodShiryaev2003} this shows that sequence $\{J_p^{(n)} \}_{n\geq 1}$ is tight. To prove the $C$-tightness, let $J_p^*$ be an accumulation point. 
 In view of \cite[Proposition~3.16, p.349]{JacodShiryaev2003} the weak convergence $J_p^{(n)} \overset{\rm d}\to J_p^*$ implies that, as $n\to\infty$,
  \beqnn
  \sum_{s\leq t} \big|J_p^{(n)}(s)-J_p^{(n)}(s-) \big|^{2+\epsilon}   \overset{\rm d}\to  \sum_{s\leq t} \big|J_p^*(s)-J_p^*(s-) \big|^{2+\epsilon} \quad  \mbox{in $\mathbf{D}(\mathbb{R}_+;\mathbb{R})$}.
  \eeqnn
 For each $t\geq 0$, it hence follows from Fatou's lemma along with Assumption \ref{main.Condition.02}, Fubini's theorem and  Lemma~\ref{V2p} that
  \beqnn
 \mathbf{E}\Big[  \sum_{s\leq t} \big|J_p^*(s)-J_p^*(s-) \big|^{2+\epsilon} \Big]  \ar=\ar  \mathbf{E}\Big[\liminf_{n\to\infty} \sum_{s\leq t} \big|J_p^{(n)}(s)-J_p^{(n)}(s-) \big|^{2+\epsilon}  \Big] \cr
 \ar\leq\ar  \liminf_{n\to\infty}\mathbf{E}\Big[ \sum_{s\leq t} \big|J_p^{(n)}(s)-J_p^{(n)}(s-) \big|^{2+\epsilon}  \Big] \cr
  \ar=\ar \liminf_{n\to\infty} \mathbf{E}\Big[ \int_0^{t} \int_{\mathbb R} \int_0^{V^{(n)}(s-)} \Big(\frac{u}{ n^\alpha}\Big)^{2+\epsilon} N_{n,0}(n\cdot ds, du, n^{2\alpha-1}\cdot dz) \Big]\cr
  \ar\leq\ar\liminf_{n\to\infty} \frac{1}{n^{\alpha\epsilon}} \int_{\mathbb R} |u|^{2+\epsilon} \nu_n(du)\cdot \int_0^t \mathbf{E}\big[ V^{(n)}(s)\big] ds \cr 
  \ar = \ar 0.
  \eeqnn 
 This shows that $J_p^*$ is almost surely continuous. Using similar arguments as in {\bf Step 2} of the proof of Proposition~\ref{Prop.3.13}, the sequence $\{J^{(n)}_p \}_{n\geq 1}$ thus converge weakly in $\mathbf{D}(\mathbb{R}_+;\mathbb{R})$ to the continuous martingale
  \beqnn
   \int_0^t \sigma_p \sqrt{V_*(s)}dW(s),\quad t\geq 0,
  \eeqnn
  with $W$ being a standard Brownian motion. 
  
 Hence, the sequence $\{ P^{(n)} \}_{n\geq 1}$ is $C$-tight and so is the sequence $\{(P^{(n)},P^{(n)}_0,I^{(n)}_p,J^{(n)}_p)\}_{n\geq 1}$. 
  Passing both sides of equation (\ref{eqn.Price01}) to the corresponding limits, we see that any accumulation point $P_*$ satisfies the stochastic equation (\ref{SL.Price}).
  
 It remains to verify that the two Brownian motions $B$ and $W$ are independent, i.e.~that $\langle B,W \rangle_t \overset{\rm a.s.}=0$ for any $t\geq 0$. To this end, we recall the martingale $M^{(n)}$ defined in (\ref{MaringaleMn}). 
 Since that $N_1(ds, dz)= N_0(ds, \mathbb{R}, dz)$,
 the covariation of the two martingales $M^{(n)}$ and $J^{(n)}_p$ is given by 
 \beqnn
 \big[ M^{(n)}, J^{(n)}_p  \big]_t
 \ar=\ar \int_0^{t} \int_{\mathbb R} \int_0^{V^{(n)}(s-)} \frac{u}{ n^{2\alpha}} N_{n,0}(n\cdot ds, du, n^{2\alpha-1}\cdot dz)\cr
 \ar=\ar  \int_\mathbb{R} u\nu_n(du) \int_0^t V^{(n)}(s)ds  \cr
 \ar\ar + \int_0^{t} \int_{\mathbb R} \int_0^{V^{(n)}(s-)} \frac{u}{ n^{2\alpha}} \widetilde{N}_{n,0}(n\cdot ds, du, n^{2\alpha-1}\cdot dz)\cr
 \ar=\ar  \int_\mathbb{R} u\nu_n(du) \int_0^t V^{(n)}(s)ds + \frac{J^{(n)}_p (t)}{n^\alpha}, 
 \eeqnn
  which vanishes weakly as $n\to\infty$ because of Assumption~\ref{main.Condition.02} and the fact that $(V^{(n)}, J^{(n)}_p)\overset{\rm d}\to (V_*,J^*_p)$. 
  This along with Theorem~6.26 in \cite[p.384]{JacodShiryaev2003} yields that  
  \beqnn
  [M_*, J_p^*]_t 
  =\int_0^t V_*(s)d \langle B,W \rangle_t 
  {\blue \overset{\rm d}=} \lim_{n\to\infty}  \big[ M^{(n)}, J^{(n)}_p  \big]_t {\blue \overset{\rm d}=}  0 ,
  \eeqnn 
  which along with the two facts that $V_* \neq   0$ and $\langle B,W \rangle_t= \rho \cdot t$ for some constant $\rho \in[-1,1]$ induces that
  \beqnn 
  \langle B,W \rangle_t {\blue\underset{\rm d}{\overset{\rm a.s.}=}} 0 , \quad t\geq 0.
  \eeqnn
  \qed  
    
\subsection{Counterexamples}\label{Sec.counter}

 Our main result states that under mild assumption the sequence of rescaled volatility processes converges to the unique (in law) solution of a stochastic differential equation. 
 Assumptions \ref{main.Condition} and \ref{main.Condition.02} are satisfied if, for instance, we choose $V_{n,0} = V_*(0)\cdot n^{2\alpha-1}$, $ \mu_n=a\cdot n^{\alpha-1}$, $\zeta_n=(1-b\cdot n^{-\alpha})^+$, $P_{n,0}=P_*(0)\cdot n^\alpha$, and $\nu_n$ being normal distributions with mean $b_p\cdot n^{-\alpha}$ and variance $\sigma_p^2$.

 We emphasize that our proof requires the  $C$-tightness of the sequence $\{J^{(n)}\}_{n \geq 1}$ to identify its limit by identifying the limit of the sequence $\{\mathcal{I}_{J^{(n)}}\}_{n \geq 1}$. Without the convergence of the sequence $\{J^{(n)}\}_{n \geq 1}$ it is not clear to us that the sequence $\{\mathcal{I}_{J^{(n)}}\}_{n \geq 1}$ converges. 
Even if the sequence of integrated processes converges, we can in general not expect to identify the weak limit of the sequence $\{J^{(n)}\}_{n \geq 1}$ by identifying the weak limit of the sequence $\{\mathcal{I}_{J^{(n)}}\}_{n \geq 1}$ as shown by the following example. The main issue is the convergence of the initial state of the volatility process that cannot always be inferred from the convergence of the integrated processes.  

%
%
   
  \begin{example}
  Let us assume that
  \beqnn
   \Lambda_n(t) =  V_{n,0} \cdot \phi(t),\quad t\geq 0.
  \eeqnn
  In this case, we have that $\Lambda_n(t) + R_n*\Lambda_n (t) =V_{n,0}\cdot R_n(t)/\zeta_n $ and 
  \beqnn
  V_n(t) \ar=\ar  \frac{V_{n,0}}{\zeta_n} R_n(t) 
  + \mu_n + \mu_n\int_0^t R_n(s)ds
     + \int_0^t \int_0^{V_n(s-)} \zeta_n   R_n (t-s) \widetilde{N}_{n,1}(ds,dz).
  \eeqnn
  We now distinguish two cases.
  
  \begin{enumerate}
  	\item[(1)] If $\alpha \in(1/2,1)$, we choose $V_{n,0}=V(0) \cdot n^\alpha$ and consider the  rescaled processes
  	\beqnn
  	V^{(n)}(t):= \frac{V_n(nt)}{n^{2\alpha-1}} .
  	\eeqnn
  	In  this case, the integrated process satisfies
  	\beqnn
  	\mathcal{I}_{V^{(n)}}(t)\ar=\ar  \frac{V(0)}{\zeta_n} \int_0^tn^{1-\alpha}R_n(ns)ds + \frac{\mu_n\cdot t}{n^{2\alpha-1}} + \frac{\mu_n }{n^{2\alpha-1}}  \int_0^t \int_0^{ns}R_n(r)drds\cr
  	\ar \ar + \int_0^t \int_0^s\int_0^{V^{(n)}(r-)} \frac{\zeta_n}{n^{2\alpha-1}}  R_n \big(n(s-r)\big) \widetilde{N}_{n,1}(n\cdot dr,n^{2\alpha-1}\cdot dz)ds. 
  	\eeqnn
  	Based on our preceding analysis, it is not difficult to see that $\mathcal{I}_{V^{(n)}}$ converges weakly to $X$ in $\mathbf{D}(\mathbb{R}_+;\mathbb{R}_+)$ with $X$ being the unique solution of the ODE
  	\beqnn
  	X(t)= \frac{V(0)}{b} F^{\alpha,\gamma}(t) + \int_0^t \frac{a}{b}F^{\alpha,\gamma}(s)ds +
  	\int_0^t \frac{1}{b} f^{\alpha,\gamma}(t-s)B(X(s))ds ,\quad t\geq 0.
  	\eeqnn
  	The random variable $X(t)$ can we written as 
  	\beqnn
  	X(t) = \int_0^t Y(s)ds,
  	\eeqnn 
  	where the process $Y$ satisfies the SDE
  	\beqnn
  	Y(t)= \frac{V(0)}{b} f^{\alpha,\gamma}(t) +\frac{a}{b}F^{\alpha,\gamma}(t) +
  	\int_0^t \frac{1}{b} f^{\alpha,\gamma}(t-s) \sqrt{Y(s)}dB(s). 
  	\eeqnn
  	However, if $V(0) \neq 0$, then $Y(t)\to \infty $ as $t\to 0+$ and hence it is impossible to prove that $ V^{(n)} \to Y $ weakly in $\mathbf{D}(\mathbb{R}_+;\mathbb{R}_+)$.

  	\item[(2)] If $\alpha>1$, then the kernel $\phi$ is light-tailed. We choose $\sigma>0$ such that 
  	\beqnn
  	\vartheta:= \int_0^\infty t \phi(t)dt \in (1,\infty). 
  	\eeqnn 
  	Similarly as in \cite{JaissonRosenbaum2015,Xu2021},  we set $V_{n,0}=V(0) \cdot n $, $\mu_n\equiv a$ and consider the  rescaled processes
  	\beqnn
  	V^{(n)}(t):= \frac{V_n(nt)}{n} .
  	\eeqnn
  	In this case,  the integrated process satisfies
  	\beqnn
  	\mathcal{I}_{V^{(n)}}(t)\ar=\ar  \frac{V(0)}{\zeta_n} \int_0^tR_n(ns)ds + \frac{a\cdot t}{n} +  a \int_0^t \int_0^{ s}R_n(nr)drds\cr
    \ar\ar + \int_0^t \int_0^s\int_0^{V^{(n)}(r-)} \frac{\zeta_n}{n }  R_n \big(n(s-r)\big) \widetilde{N}_{n,1}(n\cdot dr,n \cdot dz)ds. 
  	\eeqnn
  	By Lemma 4.5 in \cite{JaissonRosenbaum2015}, we have $\int_0^tR_n(ns)ds\to  \frac{1}{b}\big(1-e^{-\frac{b}{\vartheta}\cdot t}\big) $ uniformly.
  	Hence  it is not difficult to see that $\mathcal{I}_{V^{(n)}}$ converges weakly to $X$ in $\mathbf{D}(\mathbb{R}_+;\mathbb{R}_+)$ with $X$ being the unique solution of the ODE
  	\beqnn
  	X(t)= \frac{V(0)}{b} \Big( 1- e^{-\frac{b}{\vartheta}\cdot t}\Big)+ \int_0^t\frac{a}{b}\big(1-e^{-\frac{b}{\vartheta}\cdot s}\big) ds +
  	\int_0^t \frac{1}{b} e^{-\frac{b}{\vartheta}\cdot (t-s)} B(X(s))ds ,\quad t\geq 0.
  	\eeqnn
  	The random variable $X(t)$ can we written as 
  	\beqnn
  	X(t) = \int_0^t Y(s)ds,
  	\eeqnn 
  	where the process $Y$ satisfies the SDE
  	\beqnn
  	Y(t)=   \frac{V(0)}{\vartheta} e^{-\frac{b}{\vartheta}\cdot t} + \frac{a}{b}\big(1-e^{-\frac{b}{\vartheta}\cdot t}\big)+
  	\int_0^t \frac{1}{\vartheta} e^{-\frac{b}{\vartheta}\cdot (t-s)} \sqrt{Y(s)}dB(s), \quad t \geq 0. 
  	\eeqnn
   Since 
  	\[
  	V^{(n)}(0)= \frac{\Lambda_{n}(0) +\mu^{(n)}}{n}\to V(0) \neq Y(0)=\frac{V(0)}{\vartheta},
  	\]
  	if $V(0) \neq 0$, we see that the sequence $\{V^{(n)}\}_{n \geq 1}$ does not converge weakly to $Y$ in $\mathbf{D}(\mathbb{R}_+;\mathbb{R}_+)$. 
  	\end{enumerate}
  
 \end{example}

We acknowledge that  this example does not apply if the limiting initial volatility equals zero. In our view a vanishing initial volatility fundamentally contradicts the idea of Hawkes processes whose sole purpose is to capture the impact of past events on the arrivals of future ones. Setting the initial volatility to zero means no events occurred prior to time zero, in which case the dynamics of past and future event arrivals is not consistent.

%% file: MittagLeffler.tex
 \section{Mittag-Leffler function}\label{App.ML}
   \setcounter{equation}{0}

   In this section, we recall some elementary properties of Mittag-Leffler function. The reader may refer to \cite{HauboldMathaiSaxena2011,Mainardi2014} for a detailed discussion of Mittag-Leffler function. 
   For two constants  $\alpha,\kappa>0$, the {\sl Mittag-Leffler function} $E_{\alpha,\kappa}$ on $\mathbb{R} $ is given by 
   \beqnn
   E_{\alpha,\kappa}(x):= \sum_{n=0}^{\infty}\frac{x^n}{\Gamma(\alpha n +\kappa)},
   \quad  x\in\mathbb{R}.
   \eeqnn
   It is locally $\alpha$-H{\"o}lder continuous.  
   For $\alpha\in (0,1)$ and a constant $\gamma>0$, we denote by $F^{\alpha,\gamma}$ and $f^{\alpha,\gamma}$ the  {\sl Mittag-Leffler distribution} and {\sl density function} on $\mathbb{R}_+$ with parameters $(\alpha,\gamma)$; they are defined by 
   \beqnn
   F^{\alpha,\gamma}(t): = 1- E_{\alpha,1}(-\gamma t^\alpha)
   \quad \mbox{and}\quad 
   f^{\alpha,\gamma}(t): =\gamma \cdot t^{\alpha-1} E_{\alpha,\alpha}(-\gamma\cdot t^\alpha),\quad t>0.
   \eeqnn
   The Laplace transform of Mittag-Leffler distribution admits the representation
   \beqnn
   \mathcal{L}_{f^{\alpha,\gamma}}(\lambda) :=  \int_0^{\infty}e^{-\lambda s}f^{\alpha,\gamma}(t)dt=\frac{\gamma}{\gamma+\lambda^{\alpha}},\ \ \lambda\geq 0.
   \eeqnn
 The asymptotic behavior of the density function $f^{\alpha,\gamma}$ near zero, respectively infinity is given by
 \beqnn
 f^{\alpha,\gamma}(t) \sim \frac{\gamma \cdot t^{\alpha-1}}{\Gamma(\alpha)} 
  \quad \mbox{as $t\to 0+$}
 \quad \mbox{and} \quad
  f^{\alpha,\gamma}(t)  \sim \frac{\alpha \cdot t^{-\alpha-1}}{\gamma\Gamma(1-\alpha)} 
  \quad \mbox{as $t\to \infty$} . 
 \eeqnn
 
 We also define the function
 \beqlb\label{eqn.RML}
 R^{\alpha,\gamma}(t):= \frac{\gamma}{\Gamma(\alpha)}\cdot t^{\alpha-1},\quad t\geq0,
 \eeqlb
 whose Laplace transform admits the representation
 \beqnn
 \mathcal{L}_{R^{\alpha,\gamma}}(\lambda) :=  \int_0^{\infty}e^{-\lambda s}R^{\alpha,\gamma}(t)dt = \gamma\cdot \lambda^{-\alpha},\quad \lambda >0
 \eeqnn
 It is easy to identify that $\mathcal{L}_{R^{\alpha,\gamma}}=
  \mathcal{L}_{f^{\alpha,\gamma}}+
  \mathcal{L}_{R^{\alpha,\gamma}}\cdot 
  \mathcal{L}_{f^{\alpha,\gamma}}$, 
 which yields that $R^{\alpha,\gamma}$ is the resolvent of $f^{\alpha,\gamma}$, i.e.
 \beqlb\label{RML}
  R^{\alpha,\gamma}(t) =f^{\alpha,\gamma}(t)+ f^{\alpha,\gamma}*R^{\alpha,\gamma}(t),\quad t> 0.
 \eeqlb
 The next proposition is a direct consequence of the variation of constants formula for linear Volterra integral equations as given in  \cite[p.36, Equation (1.2)]{GripenbergLondenStaffans1990}. 
 Its proof is omitted.
  
 \begin{proposition}\label{eqn.App}
  For a function $H \in L^1_{\rm loc}(\mathbb{R}_+;\mathbb{R}) $, the unique solution to  
  \beqnn
  X(t)= H(t) + f^{\alpha,\gamma}*X(t),\quad t\geq 0,
  \eeqnn
   admits the representation 
  \beqnn
  X(t) =H(t) +  R^{\alpha,\gamma} * H(t),\quad t\geq 0.
  \eeqnn
  	
 \end{proposition}
